\documentclass[traditabstract]{aa}
\usepackage{natbib}
\usepackage{graphicx}
\usepackage{xcolor}
\usepackage{amssymb, amsfonts}
\usepackage{textpos}
\usepackage{savesym}
\usepackage{amsmath}
\savesymbol{iint}
\usepackage{txfonts}
\restoresymbol{TXF}{iint}

\usepackage[latin1]{inputenc}
\usepackage{subfigure}
\usepackage{booktabs}
\usepackage{url}
\usepackage{hyperref}
\hypersetup{
    colorlinks,
    citecolor=black,%
    filecolor=black,%
    linkcolor=black,%
    urlcolor=black
}

\begin{document}

\title{Detectability of extrasolar moons as gravitational microlenses}

\author{Christine Liebig \and Joachim Wambsganss}

\institute{Astronomisches Rechen-Institut, Zentrum f\"{u}r Astronomie
  der Universit\"{a}t Heidelberg, M\"onchhofstra{\ss}e 12-14, 69120
  Heidelberg, Germany, \email{cliebig@ari.uni-heidelberg.de,
    jkw@ari.uni-heidelberg.de}}

\date{Received 10 December 2009 / Accepted 25 April 2010  }

\abstract{We evaluate gravitational lensing as a technique for the
  detection of extrasolar moons.  Since 2004 gravitational
  microlensing has been successfully applied as a detection method for
  extrasolar planets. In principle, the method is sensitive to masses
  as low as an Earth mass or even a fraction of it. Hence it seems
  natural to investigate the microlensing effects of moons around
  extrasolar planets. We explore the simplest conceivable triple lens
  system, containing one star, one planet and one moon. From a
  microlensing point of view, this system can be modelled as a
  particular triple with hierarchical mass ratios very different from
  unity.  Since the moon orbits the planet, the planet-moon separation
  will be small compared to the distance between planet and star. Such
  a configuration can lead to a complex interference of caustics. We
  present detectability and detection limits by comparing triple-lens
  light curves to best-fit binary light curves as caused by a
  double-lens system consisting of host star and planet -- without
  moon.  We simulate magnification patterns covering a range of mass
  and separation values using the inverse ray shooting
  technique. These patterns are processed by analysing a large number
  of light curves and fitting a binary case to each of them. A
  chi-squared criterion is used to quantify the detectability of the
  moon in a number of selected triple-lens scenarios. The results of
  our simulations indicate that it is feasible to discover extrasolar
  moons via gravitational microlensing through frequent and highly
  precise monitoring of anomalous Galactic microlensing events with
  dwarf source stars.}

\keywords{gravitational lensing -- planetary systems -- methods:
  numerical -- methods: statistical }

\titlerunning{Exomoons as gravitational lenses}
\authorrunning{C. Liebig \& J. Wambsganss}

\maketitle

\section{Introduction }
\label{sec:intro}

By now hundreds of extrasolar planets have been
detected\footnote{\url{exoplanet.eu}}.  For all we know, none of the
newly discovered extrasolar planets offers physical conditions
permitting any form of life. But the search for planets potentially
harbouring life and the search for indicators of habitability is
ongoing.  One of these indicators might be the presence of a large
natural satellite -- a moon -- which stabilises the rotation axis of
the planet and thereby the surface climate \citep{Benn2001}. It has
also been suggested that a large moon itself might be a good candidate
for offering habitable conditions \citep{Scharf2006}.  In the solar
system, most planets harbour moons. In fact, the moons in the solar
system outnumber the planets by more than an order of magnitude. No
moon has yet been detected around an extrasolar planet.

The majority of known exoplanets has been discovered through radial
velocity measurements, with the first successful finding reported by
\citet{Mayor1995}. This method is not sensitive to satellites of those
planets, because the stellar ``Doppler wobble'' is only affected by
the orbital movement of the barycentre of a planet and its satellites,
though higher-order effects could play a role eventually. Here we
consider Galactic microlensing, which has led to the discovery of
several relatively low-mass exoplanets since the first report of a
successful detection by \citet{Bond2004}, as a promising technique for
the search for exomoons.

As early as 1999, it has been suggested that extrasolar moons might be
detectable through transit observations \citep{Sartoretti1999}, either
through direct observation of lunar occultation or through transit
timing variations, as the moon and planet rotate around their common
barycentre, causing time shifts of the transit ingress and egress
(cf. also \citet{Holman2005}). In their simulations of space-based
gravitational microlensing \citet{Bennett2002} mention the possibility
of discovering extrasolar moons similar to our own Moon. Later that
year, \citet{Han2002} performed a detailed feasibility study whether
microlensing offers the potential to discover an Earth-Moon analogue,
but concluded that finite source effects would probably be too severe
to allow detections. \citet{Williams2004} published the quite original
suggestion to look for spectral signatures of Earth-sized moons in the
absorption spectra of Jupiter-sized planets. \citet{Cabrera2007}
proposed a sophisticated transit approach using ``mutual event
phenomena'', i.e. photometric variation patterns due to different
phases of occultation and light reflection of planet and
satellite. \citet{Han2008a} undertook a new qualitative study of a
number of triple-lens microlensing constellations finding
``non-negligible'' light curve signals to occur in the case of an
Earth-mass moon orbiting a 10 Earth-mass planet, ``when the
planet-moon separation is similar to or greater than the Einstein
radius of the planet''.  \citet{Lewis2008} analysed pulsar
time-of-arrival signals for lunar signatures. \citet{Kipping2009a,
  Kipping2009b} refined and extended the transit timing models of
exomoons to include transit duration variations, and
\citet{Kipping2009c} examined transit detectability of exomoons with
Kepler-class photometry and concluded that in optimal cases moon
detections down to $0.2\,M_\text{Earth}$ should be possible.

We cover here several new aspects concerning the microlensing search
for exomoons, extending the work of \citet{Han2008a}.  First, the
detectability of lunar light curve perturbations is determined with a
statistical significance test that does not need to rely on human
judgement.  Second, all parameters of the two-dimensional three-body
geometry, including the position angle of the moon with respect to the
planet-star axis, are varied.  Third, an unbiased extraction of light
curves from the selected scenarios enables a tentative prediction of
the occurrence rate of detectable lunar light curve signals.
A more detailed account of this study is available as 
\citet{Liebig2009}. 

The paper is structured as follows: 
In Section~\ref{sec:basics}, we recall the fundamental
equations of gravitational microlensing relevant to our work. 
Our method for quantifying the detection
rates for extrasolar moons in selected lensing scenarios is presented in
Section~\ref{sec:method}. In Section~\ref{sec:simulatedscenarios}, we
discuss the astrophysical implications of the input parameters of the
simulations. Our results are presented in Section~\ref{sec:results},
together with a first interpretation and a discussion of potential
problems of our method.

\section{Basics of gravitational microlensing}
\label{sec:basics}

The deflection of light by massive bodies is a consequence of the
theory of general relativity \citep{Einstein1916} and has been
experimentally verified since 1919 \citep{Dyson1920}, see
\citet{Paczynski1996} for an introduction to the field or 
\citet{Schneider2006} for a comprehensive review.

The typical scale of angular separations in gravitational lensing is
the \emph{Einstein radius} \ensuremath{\theta_{\textit{E}}}, the
angular radius of the ring of formally infinite image magnification
that appears when a source at a distance $D_S$, a lens of mass $M$ at
a distance $D_L$, and the observer are perfectly aligned:
\begin{align}\label{eq:Einsteinradius}
  \theta_{E}= \sqrt{\frac{4GM}{c^2}\frac{D_{LS}}{D_LD_S}},
\end{align} 
where $G$ denotes the gravitational constant, $c$ the speed of
light. $D_{LS}$ is the distance between lens plane and source plane;
in the non-cosmological distance scale of our galaxy $D_{LS} = D_S -
D_L$ holds true. With the Einstein radius also comes the
characteristic time scale of transient gravitational lensing events,
the \emph{Einstein time} $t_E=D_L\theta_E/v_\perp$ with the transverse
velocity $v_\perp$ of the lens relative to the source.

Here we focus on the triple lens case with host star ($S$), planet
($P$) and moon ($M$), see Figure~\ref{fig:mp-parameters} for
illustration. The lens equation can be expressed using complex
coordinates, where $\eta$ shall denote the angular source position and
$\xi$ the image positions, cf. choice of notation in \citet{Witt1990}
and \citet{Gaudi1998}. $\xi_i$ stands for the angular position of the
lensing body $i$. $q_{ij}$ is the mass ratio between lenses $i$ and
$j$ ($q_{ij}=\frac{M_i}{M_{j}}$). The lens equation gets the following
form, if the primary lens, the host star $S$, has unit mass and is
placed in the origin of the lens plane,
\begin{align}\label{eq:triplelensequation}  
  \eta = \xi - \frac{1}{\overline{\xi}} -
  \frac{q_{\text{PS}}}{\overline{\xi} -
    \overline{\xi_{\text{P}}}} -
  \frac{q_{\text{MS}}}{\overline{\xi} -
    \overline{\xi_{\text{M}}}}.
\end{align}
This is a mapping from the lens plane to the source plane, which maps
the images of a source star to its actual position in the source
plane.  As pointed out in \citet{Rhie1997}, and explicitly calculated
in \citet{Rhie2002}, the triple-lens equation is a tenth-order
polynomial equation in $\xi$.

In microlensing, the images cannot be resolved. Detectable is only the
transient change in magnitude of the source star, when lens and source
star are in relative motion to each other. The total magnification
$\mu$ of the base flux is obtained as the inverse of the determinant
of the Jacobian of the mapping equation~\eqref{eq:triplelensequation},
\begin{align}\label{eq:magnification}
  \mu = \frac{1}{\det J(\xi)} \text{, with } \det J(\xi)=
  1-\frac{\partial\eta}{\partial \overline{\xi}}\frac{\overline{\partial
      \eta}}{\partial \overline{\xi}}.
\end{align}
Gravitational lensing changes the apparent solid angle of a source,
not the surface brightness. The magnification $\mu$ is the ratio of
the total solid angle of the images and the apparent solid angle of
the unlensed source.

For a status report of the past, present and prospective future of
planet searching via Galactic microlensing the recent white papers of
2008/2009 are a good source of reference 
\citep{Beaulieu2008,Dominik2008b,Bennett2009b,Gaudi2009}.

\section{Method} 
\label{sec:method}

\begin{figure}[t]
  \centering
  \resizebox{\columnwidth}{!}{\input{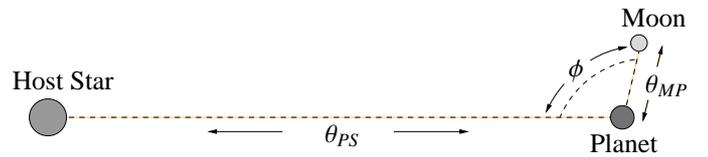}}
  \caption{Geometry of our triple-lens scenario, not to scale.  Five
    parameters have to be fixed: The mass ratios $q_{PS} =
    M_{\text{Planet}}/M_{\text{Star}}$ and $q_{MP} =
    M_{\text{Moon}}/M_{\text{Planet}}$, the angular separations in the
    lens plane $\theta_{PS}$ and $\theta_{MP}$ and the position angle
    of the moon $\phi$. These parameters define uniquely the relative
    projected positions of the three
    bodies.} \label{fig:mp-parameters}
\end{figure}

The simplest gravitational lens system incorporating an extrasolar
moon is a \emph{triple-lens} system consisting of the lensing star, a
planet and a moon in orbit around that planet, as sketched in
Figure~\ref{fig:mp-parameters}. Most likely, lunar effects will first
show up as noticeable irregularities in light curves that have been
initially observed and classified as light curves with planetary
signatures.  To measure the detectability of a given triple-lens
system among binary lenses, we have to determine whether the provided
triple-lens light curve differs statistically significantly from
binary-lens light curves.

To clarify the terminology: We investigate light curves that show a
deviation from the single-lens case due to the presence of the
planetary caustic, i.e. which would be modelled as a star-plus-planet
system in a first approximation.  We call lunar \emph{detectability}
the fraction of those light curves which display a significant
deviation from a star-plus-planet lens model due to the presence of
the moon.  We measure the difference between a given triple-lens light
curve (which is taken to be the ``true'' underlying light curve of the
event) and its best-fit binary-lens counterpart. The best-fit
binary-lens light curve is found by a least-square fit. We then employ
$\chi^2$-statistics to see whether the triple-lens light curve could
be explained as a normal fluctuation within the error boundaries of
the binary-lens light curve. If this is not the case, the moon is
considered \emph{detectable}.

Detection and characterisation are two separate problems in the search
for extrasolar planets or moons, though, and we do not make statements
about the latter. When characterising an observed light curve with
clear deviations from the binary model, it is still possible that
ambiguous solutions -- lunar and non-lunar -- arise. This does not
reflect on our results. We simply give the fraction of triple-lens
light curves significantly deviating from binary-lens light curves,
without exploring whether they can be uniquely characterised.

A qualitative impression of the lunar influence on the caustic
structure can be gained from the magnification patterns in
Figure~\ref{fig:orbit}.  Extracted example light curves are presented
in Figure~\ref{fig:orbit-lc2}. Going beyond the qualitative picture,
in this paper we quantify the detectability of an extrasolar moon in
gravitational microlensing light curves in selected scenarios.

\setlength{\unitlength}{0.3\textwidth}
\begin{figure*}[p]
\begin{picture}(3,4.47)(0,0.03)
\put(0,0){\includegraphics[width=\textwidth]{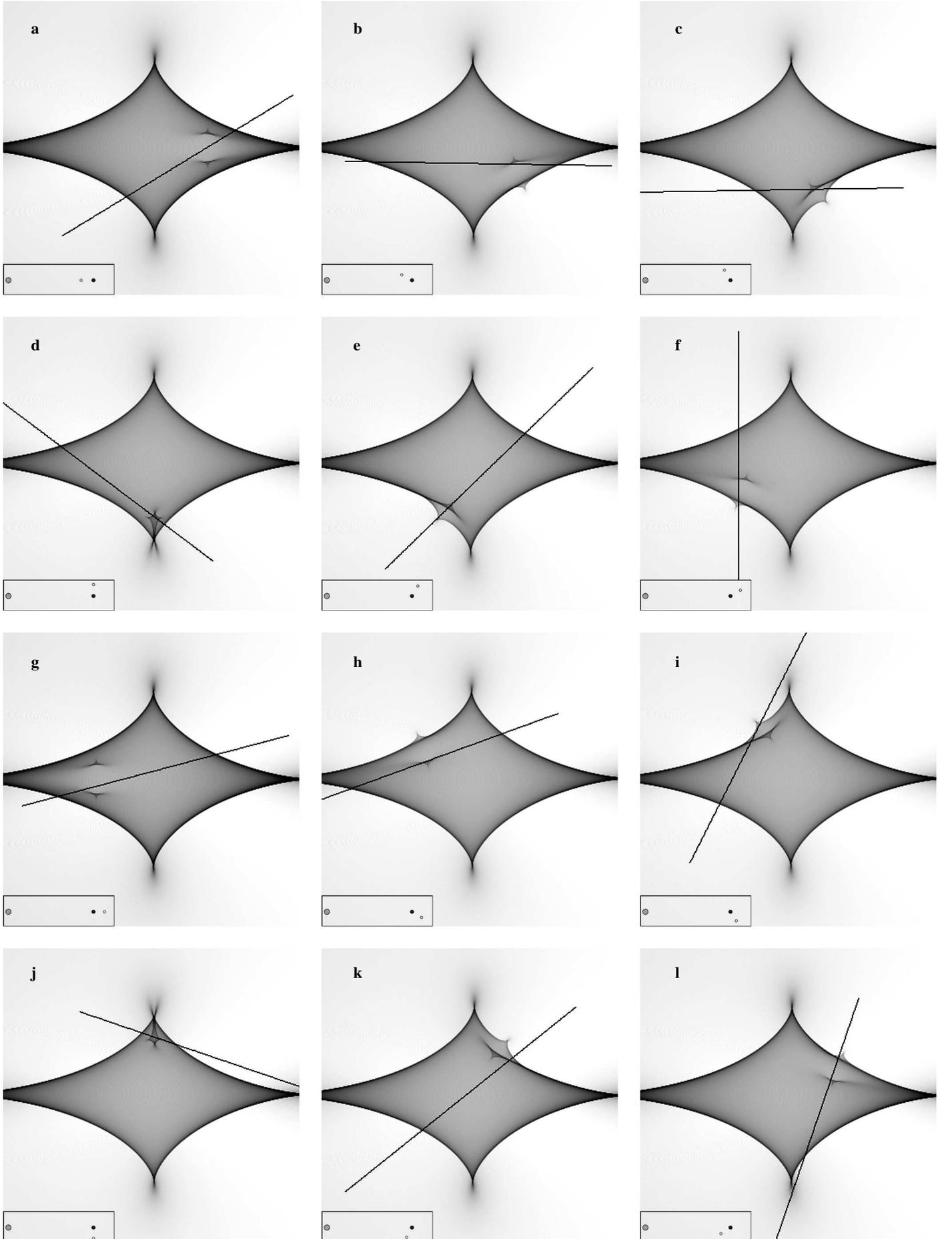}}
\put(0.1,4.35){\textbf{a}}
\put(1.25,4.35){\textbf{b}}
\put(2.4,4.35){\textbf{c}}
\put(0.1,3.21){\textbf{d}}
\put(1.25,3.21){\textbf{e}}
\put(2.4,3.21){\textbf{f}}
\put(0.1,2.07){\textbf{g}}
\put(1.25,2.07){\textbf{h}}
\put(2.4,2.07){\textbf{i}}
\put(0.1,0.95){\textbf{j}}
\put(1.25,0.95){\textbf{k}}
\put(2.4,0.95){\textbf{l}}
\end{picture}
\caption{Details of the analysed triple-lens magnification maps for an
  example mass/separation scenario with $q_{PS}= 10^{-3}$,
  $q_{MP}=10^{-2}$, $\theta_{PS}=1.3\,\ensuremath{\theta_{\textit{E}}}$ and
  $\theta_{MP}=1.0\,\ensuremath{\theta_{\textit{E}}}^P$ (the standard
  scenario as summarised in Table~\ref{tab:standardsummary}), showing
  the caustic configurations for twelve different lunar positions
  completing a full circular orbit in steps of $30\degr$. 
  The relative positions of star, planet and moon
  are sketched in the lower left of each panel (not to scale). 
  The side
  lengths of the individual frames are
  $0.1\,\ensuremath{\theta_{\textit{E}}}$. A darker shade of grey
  corresponds to a higher magnification. The straight black lines mark
  source trajectories, the resulting light curves are displayed in
  Figure~\ref{fig:orbit-lc2}.}
  \label{fig:orbit}
\end{figure*}

\begin{figure*}[p]
\begin{picture}(3,4.47)(0,0.03)
\put(0.1,0.1){\resizebox{0.98\textwidth}{!}{\input{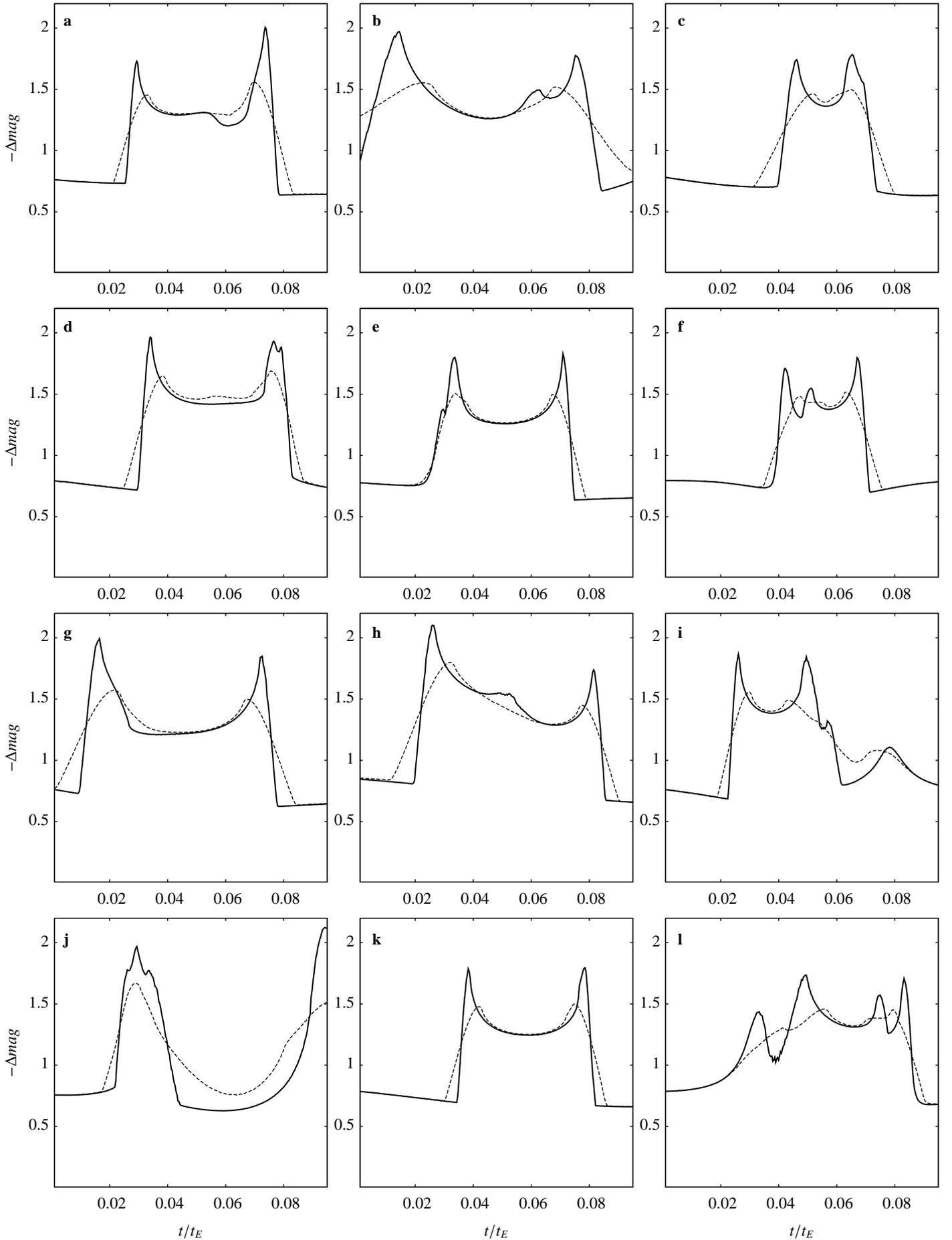}}}
\put(0.21,4.37){\textbf{a}}
\put(1.31,4.37){\textbf{b}}
\put(2.4,4.37){\textbf{c}}
\put(0.21,3.27){\textbf{d}}
\put(1.31,3.27){\textbf{e}}
\put(2.4,3.27){\textbf{f}}
\put(0.21,2.17){\textbf{g}}
\put(1.31,2.17){\textbf{h}}
\put(2.4,2.17){\textbf{i}}
\put(0.21,1.07){\textbf{j}}
\put(1.31,1.07){\textbf{k}}
\put(2.4,1.07){\textbf{l}}
\end{picture}
\caption{Sample of triple-lens light curves corresponding to the
  source trajectories depicted in Figure~\ref{fig:orbit}. The solid
  light curves were extracted with an assumed solar source size,
  $R_\text{Source} = R\ensuremath{_{\sun}}$; the thin, dashed lines
show the same light curves with $R_\text{Source} =
3\,R\ensuremath{_{\sun}}$.  The magnification scale is given in
negative magnitude difference ($-\Delta mag$), i.e. the unmagnified
baseline flux of the source is 0. The time scale in units of the
Einstein time $t_E$, assuming a uniform relative motion of source and
lens.}
  \label{fig:orbit-lc2}
\end{figure*}

\subsection{Ray shooting }
\label{sec:rs}

The triple-lens equation~\eqref{eq:triplelensequation} is analytically
solvable, but \citet{Han2002} pointed out that numerical noise in the
polynomial coefficients caused by limited computer precision was too
high ($\sim10^{-15}$) when solving the polynomial numerically for the
very small mass ratios of moon and star ($\sim10^{-5}$). To avoid
this, we employ the inverse ray-shooting technique, which has the
further advantage of being able to account for finite source sizes and
non-uniform source brightness profiles more easily. It also gives us
the option of incorporating additional lenses (further planets or
moons) without increasing the complexity of the calculations.  This
technique was developed by \citet{Schneider1986}, \citet{Kayser1986}
and \citet{Wambsganss1990,Wambsganss1999a} and already applied to
planetary microlensing in \citet{Wambsganss1997}.

\begin{figure}[tb]
  \centering 
  \includegraphics[width=\columnwidth]{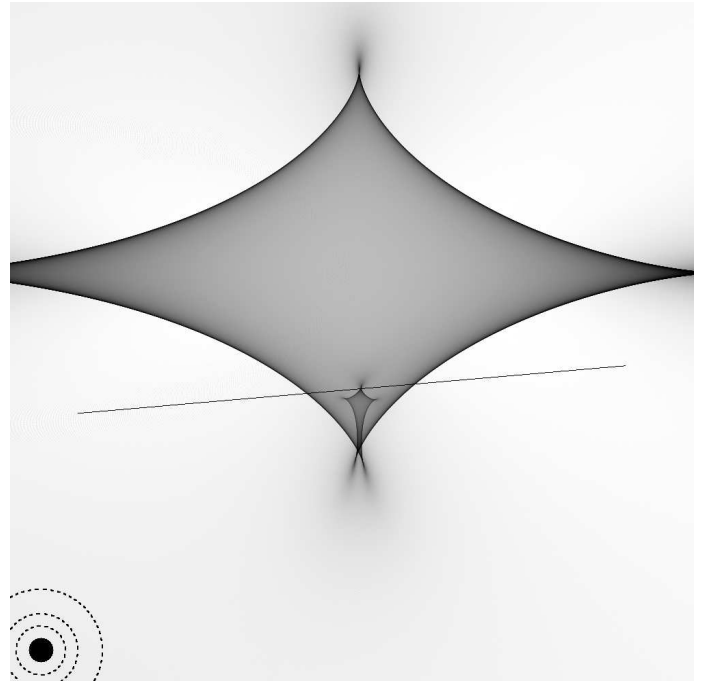}
  \caption{Magnification pattern displaying the planetary caustic of a
    triple-lens system with mass ratios $q_{PS} = 10^{-3}$ and $q_{MP}
    = 10^{-2}$, corresponding, e.g., to an M-dwarf, a Saturn-mass
    planet and an Earth-mass satellite of that planet.  Darker shades
    correspond to areas of higher magnification.  The third body shows
    its influence as the small perturbation on the bottom cusp of the
    planetary caustic.  The position angle of the moon is
    $\phi=90\degr$, i.e. in the orientation of this map the moon is
    located above the planet. The complete set of parameters is given
    in Table~\ref{tab:standardsummary}. The side length of the pattern
    is $0.1\,\ensuremath{\theta_{\textit{E}}}$ in stellar Einstein
    radii. A source with radius $R\ensuremath{_{\sun}}$
    (2$R\ensuremath{_{\sun}}$, 3$R\ensuremath{_{\sun}}$,
    5$R\ensuremath{_{\sun}}$) is indicated as a black disc (dashed
    circles) in the lower left. Taking the straight line that cuts the
    lunar perturbation of the caustic as the source trajectory of a
    solar sized source results in the light curve shown in
    Figure~\ref{fig:fitnofit} (solid line).  Depending on the assumed
    source size, different light curves are obtained, see also
    Figure~\ref{fig:lc-sourcesizes}\subref{fig:2solarsizedsource} to
    \subref{fig:5solarsizedsource}.}
  \label{fig:track-649}
\end{figure}

Inverse ray-shooting means that light rays are traced from the
observer back to the source plane. This is equivalent to tracing rays
from the background source star to the observer plane. The influence
of all masses in the lens plane on the light path is calculated.  In
the thin lens approximation, the deflection angle is just the sum of
the deflection angles of every single lens. After deflection, all
light rays are collected in pixel bins of the source plane. Thus a
magnification pattern is produced (e.g. Figure~\ref{fig:track-649}).
Lengths in this map can be translated to angular separations or,
assuming a constant relative velocity between source and lens, to time
intervals. The number of collected light rays per pixel is
proportional to the magnification $\mu$ of a background source with
pixel size at the respective position in the source plane. The
resolution in magnification of these numerically produced patterns is
finite and depends on the total number of rays shot.

\subsection{Light curve simulation}
\label{sec:LCsimulation}

\begin{figure}[tb]
  \centering
  \includegraphics[width=\columnwidth]{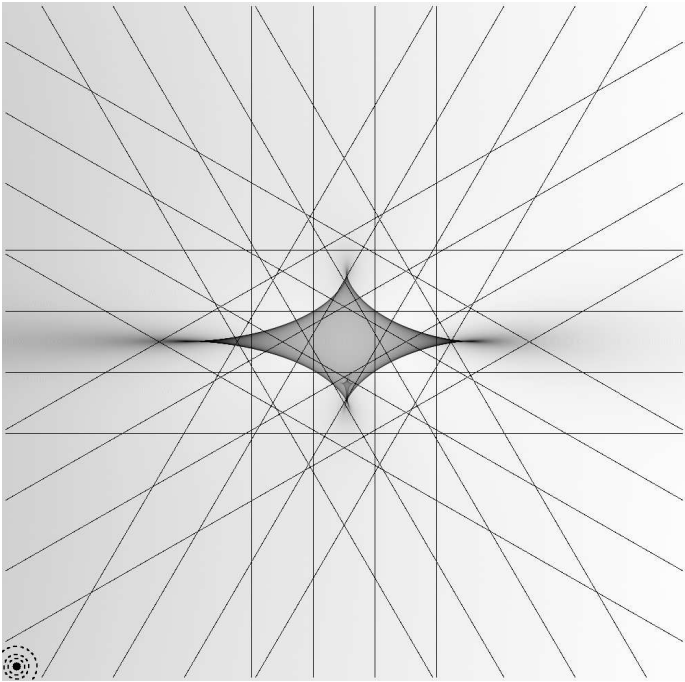}
  \caption{Illustration of how the magnification pattern of a given
    star-planet-moon configuration is evaluated. In order to have an
    unbiased statistical sample, the source trajectories are chosen
    independently of the lunar caustic features, though all light
    curves are required to pass through or close to the planetary
    caustic. This is realised through generating a grid of source
    trajectories that is only oriented at the planetary caustic. In
    the actual evaluation a 10 times denser grid than shown here
    was used.  
    The magnification map has to be
    substantially larger than the caustic to ensure an adequate
    baseline for the light curve fitting. The side length of this 
    pattern is $0.3\,\ensuremath{\theta_{\textit{E}}}$, the position
    angle of the moon $\phi=90\degr$, all other parameters are chosen
    as listed in Table~\ref{tab:standardsummary}.  Source sizes of
    $1\,R\ensuremath{_{\sun}}$, 
    $2\,R\ensuremath{_{\sun}}$, 
    $3\,R\ensuremath{_{\sun}}$, and
	$5\,R\ensuremath{_{\sun}}$ are
    indicated in the lower left.}
  \label{fig:lightcurvegrid}
\end{figure}

We obtain simulated microlensing light curves with potential traces of
a moon by producing magnification maps (Figure~\ref{fig:track-649}) of
triple-lens scenarios with mass ratios very different from unity.  A
light curve is then obtained as a one-dimensional cut through the
magnification pattern, convolved with the luminosity profile of a
star.  Only a few more assumptions are necessary to simulate
realistic, in principle observable, light curves: angular source size,
relative motion of lens and source, and lens mass.  As a first
approximation to the surface brightness profiles of stars we use a
profile with radius $R_{\text{Source}}$ and constant surface
brightness (ignoring limb-darkening effects). The light curve is
sampled at equidistant intervals. In Section~\ref{sec:sampling} we
discuss physical implications of the sampling frequency.  In order to
be able to make robust statistical statements about the detectability
of a moon in a chosen setting, i.e. a certain magnification pattern,
we analyse a grid of source trajectories that delivers about two
hundred light curves, see also Figure~\ref{fig:lightcurvegrid}. To
have an unbiased sample, the grid is chosen independently of lunar
caustic features, but all trajectories are required to show pronounced
deviations from the single-lens light curve due to the planet.

\subsection{Fitting}
\label{sec:fitting}

\begin{figure}[tb]
  \centering
\subfigure[Identical source trajectory.]
{\resizebox{1.05\columnwidth}{!}{\input{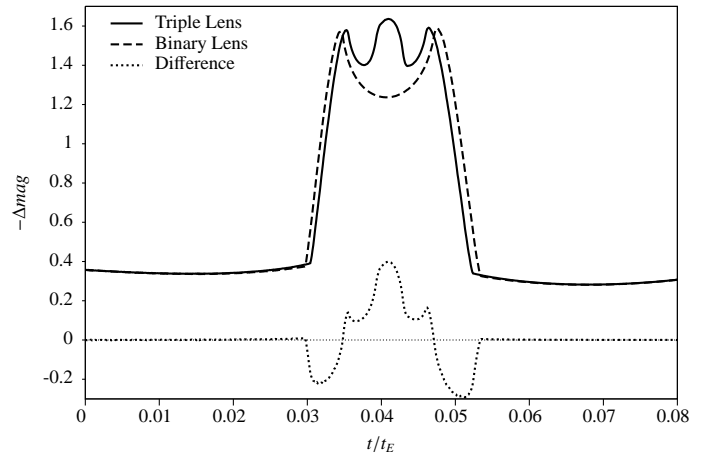}}
  \label{fig:nofit}}
\subfigure[Best-fit source trajectory.] 
{\resizebox{1.05\columnwidth}{!}{\input{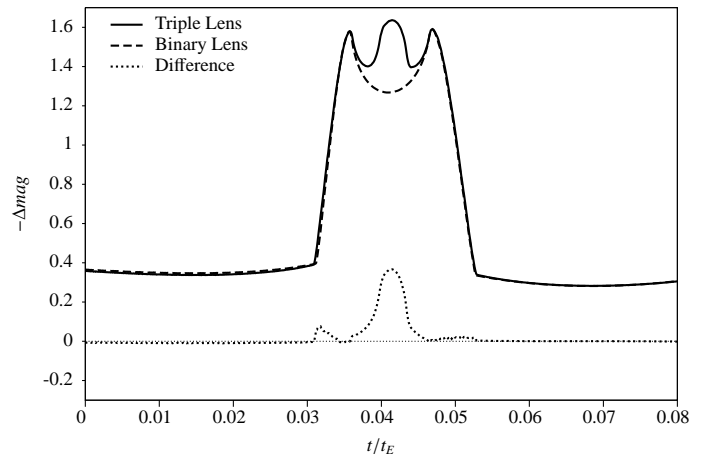}}
  \label{fig:fit}}%
\caption[]{Triple-lens light curve (solid line) extracted from the
  magnification pattern in Figure~\ref{fig:track-649}, compared with a
  light curve extracted from the corresponding binary-lens
  magnification map (dashed line), with no third body and the planet
  mass increased by the previous moon mass. The difference of the two
  light curves is plotted as the dotted line. The central peak of the
  triple-lens light curve is caused by lunar caustic perturbations,
  that cannot easily be reproduced with a binary lens. In
  \subref{fig:nofit}, identical source trajectory parameters are used
  to extract the binary-lens light curve and large residuals remain
  that can be avoided with a different source trajectory. In
  \subref{fig:fit}, the best-fit binary-lens light curve is shown
  (dashed) in comparison with the triple-lens light curve
  (solid). Here, the residuals can be attributed to the moon. If the
  deviation is significant, the moon is \emph{detectable} in the
  triple-lens light curve.  A source with solar radius at a distance
  of 8\,kpc is assumed.}
  \label{fig:fitnofit} 
\end{figure}

For each triple-lens scenario, we produce an additional magnification
pattern of a corresponding binary-lens system, where the mass of the
moon is added to the planetary mass.  As a first approximation, one
can compare two light curves with identical source track parameters,
see Figure \ref{fig:nofit}. Numerically big differences can occur
without a significant topological difference, because the additional
third body not only introduces additional caustic features, it also
slightly moves the location of the double-lens caustics.
Since the perturbation introduced by the moon is small, however, we
have good starting conditions using the light curve from the binary
magnification pattern that corresponds to the triple-lens
magnification pattern. Keeping the planet-star separation fixed and
changing the planetary mass to the sum of the masses of planet and
moon, we are likely to be close to the global best-fit values.  In
our simplified approach, we only vary the
source trajectory to search for the best-fitting binary-lens
light curve, which already yields satisfactory results, as can be
seen in Figure~\ref{fig:fit}. 
For the fitting procedure we use the simple, straight-forward and
robust least-square method.

\subsection{Light curve comparison}
\label{sec:curvecomparison}

We want to use the properties of the $\chi^2$-distribution for a test
of significance of deviation between the two simulated light curves of
the triple lens and the corresponding binary lens.

Most commonly, the $\chi^2$-distribution is used to test the
goodness-of-fit of a model to experimental data.  To compare two
simulated curves instead, as in our case, one possible approach is to
randomly generate artificial data around one of them and then
consecutively fit the two theoretical curves to the artificial data
with some free parameters. Two `$\chi^2$-values' $Q^2_1$ and $Q^2_2$
will result. Their difference, $\Delta Q^2 = Q^2_1 - Q^2_2$ can then
be calculated and a threshold value $\Delta Q^2_\text{thresh}$ is
chosen.  The condition for reported detectability of the deviation is
then $\Delta Q^2 > \Delta Q^2_\text{thresh}$.  A detailed description
of the algorithm and its application to microlensing was presented by
\citet{Gaudi2000}.

We decided to use a different approach here: 
In particular,
we were led to consider other possible techniques by the wish to avoid
``feeding'' our knowledge of the data distribution to a random number
generator in order to get an unbiased and random $\chi^2$-distributed
sample of $Q^2$, when at the same time we have all the necessary
information to calculate a much more representative $\chi^2$-value.
We have developed a method to quantify significance of deviations
between two theoretical functions that avoids the steps of data
simulation and subsequent fitting.

Put simply, we calculate the mean $\langle Q^2\rangle$ of all
`$\chi^2$-values' that would result, if data with a Gaussian scatter
drawn from a triple-lens light curve is fitted with a
binary light curve. If the triple-lens case and the best-fit
binary-lens case are indistinguishable, then $Q^2$ will be
$\chi^2$-distributed and $\langle Q^2\rangle$ will lie at or very
close to $\langle \chi^2\rangle = \textit{number of degrees of
  freedom}$, the mean of the $\chi^2$ probability density function. If
the triple-lens light curve and the binary-lens light curve differ
significantly, then $Q^2$ is not $\chi^2$-distributed and $\langle
Q^2\rangle$ in particular will lie outside the expected range for
$\chi^2$-distributed random variables. In the latter case, the moon is
detectable. This approach is presented in mathematical
detail in Appendix \ref{sec:significance}.

\section{Choice of scenarios}
\label{sec:simulatedscenarios}

This section presents the assumptions that we use for our simulations.
We discuss the astrophysical parameter space that is available for
simulations of a microlensing system consisting of star, planet and
moon. By choosing the most probable or most pragmatic value for each
of the parameters, we create a standard scenario, summarised in
Table~\ref{tab:standardsummary} and visualised in
  Figure~\ref{fig:lensingzones}, that all other parameters are
compared against during the analysis.

The observational search for microlensing events caused by extrasolar
planets is carried out towards the Galactic bulge, currently limited
to the fields of the wide-field surveys OGLE \citep{Udalski1992} and
MOA \citep{Muraki1999}. This leads to some natural assumptions for the
involved quantities. The source stars are typically close to the
centre of the Galactic bulge at a distance of $D_S=8\,\text{kpc}$,
where the surface density of stars is very high.  For the distance to
the lens plane we adopt a value of $D_L = 6\,\text{kpc}$.
We assume our primary lens mass to be an M-dwarf star with a mass of
$M_{\text{Star}}=0.3M\ensuremath{_{\sun}}$, because that is the most
abundant type, cf.  Figure~5 of \citet{Dominik2008b}. Direct
  lens mass determination is only possible if additional observables,
  such as parallax, can be measured, cf. \citet{Gould2009c}.
We derive the corresponding Einstein radius of
$\ensuremath{\theta_{\textit{E}}}(D_S, D_L, M_{\text{Star}}) =
0.32\,$milliarcseconds.

Five parameters describe our lens configuration, cf.
Figure~\ref{fig:mp-parameters}: They are the mass ratios between
planet and star $q_{PS}$ and between moon and planet $q_{MP}$, the
angular separations $\theta_{PS}$ between planet and star and
$\theta_{MP}$ between moon and planet, and as the last parameter
$\phi$, the position angle of the moon with respect to the planet-star
axis.  These parameters are barely constrained by the physics of a
three-body system, even if we do require mass ratios very different
from unity and separations that allow for stable orbits. 

The apparent source size $R_{\text{Source}}$ and the sampling rate
$f_\text{sampled}$ affect the shape of the simulated light curves. We
have to define an observational uncertainty $\sigma$ for the
significance test in the final analysis of the light curves.

\subsection{Mass ratio of planet and star $q_{PS}$}
\label{sec:q_PS}

We are interested in planets (and not binary stars). Accordingly, we
want a small value for the mass ratio of planet and star $q_{PS} =
\frac{M_{\text{Planet}}}{M_{\text{Star}}}$. Our standard value will be
$10^{-3}$ which is the mass ratio of Jupiter and Sun, or a Saturn-mass
planet around an M-dwarf of $0.3\,M\ensuremath{_{\sun}}$. At a given
projected separation between star and planet, this mass ratio
determines the size of the planetary caustic, with a larger $q_{PS}$
leading to a larger caustic. $q_{PS}$ is varied to also examine
scenarios with mass ratios corresponding to a Jovian mass around an
M-dwarf and to a Saturn mass around the Sun.
Summarised: For the mass ratio between planet and star we use the
three values $q_{PS} = 3.3\times10^{-3}, 10^{-3}, 3\times10^{-4}$.

\subsection{Mass ratio of moon and planet $q_{MP}$}
\label{sec:q_MP}
\begin{figure}[tb]
  \centering
\subfigure[$q_{MP}=10^{-3}$]{  
\includegraphics[width=0.3\columnwidth]{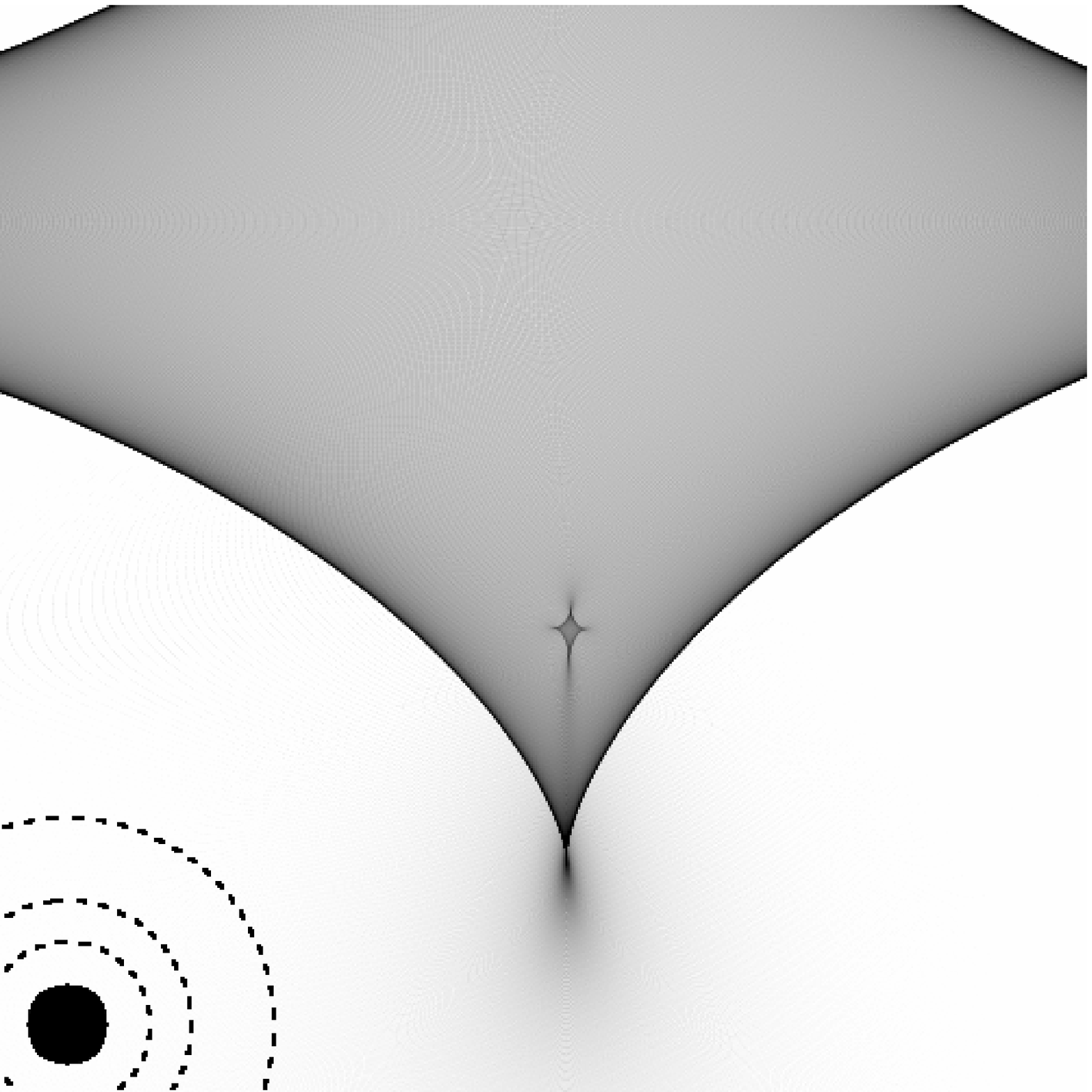}
\label{fig:sub1-caustic-q_MP}
}
\subfigure[$q_{MP}=10^{-2}$]{
\includegraphics[width=0.3\columnwidth]{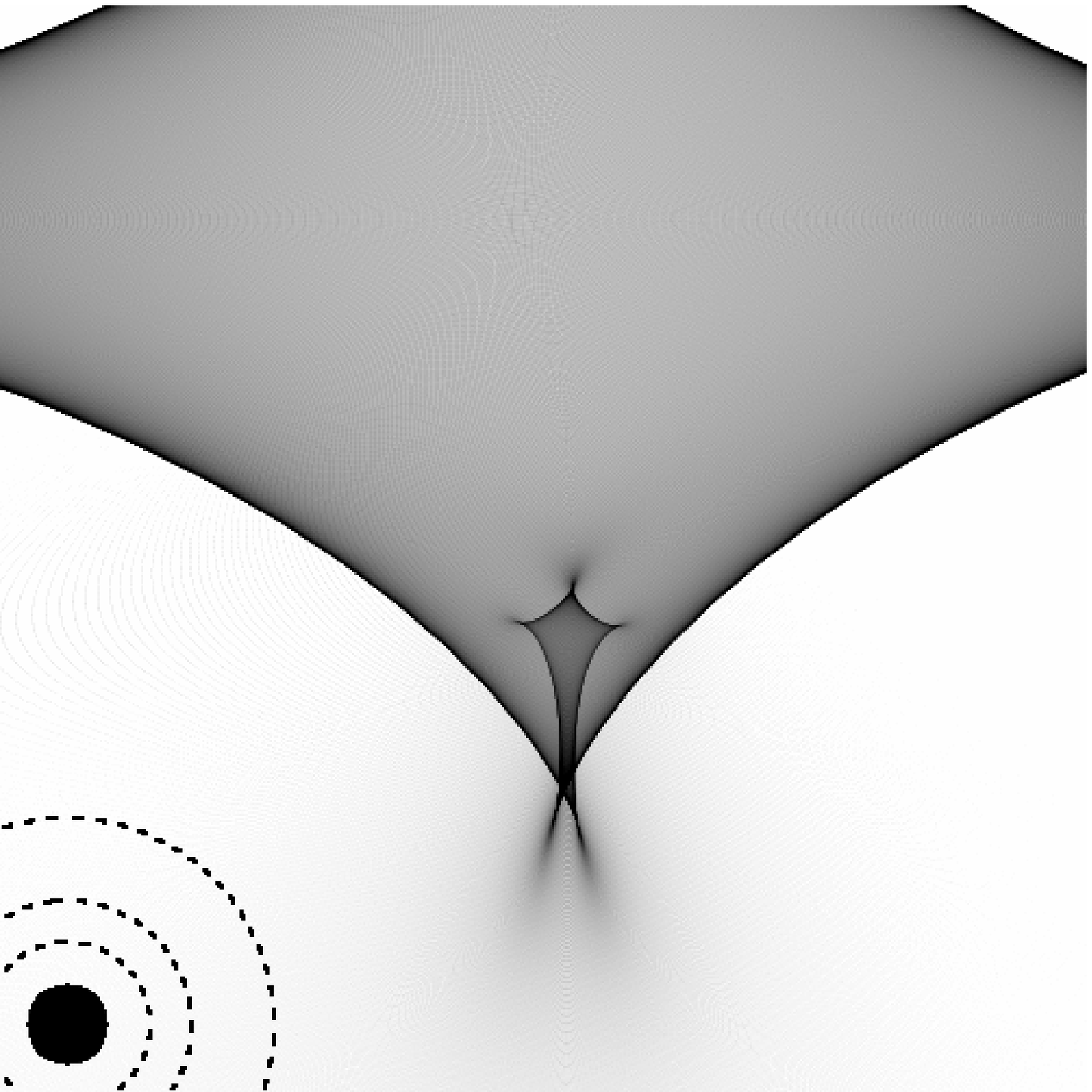}
\label{fig:sub2-caustic-q_MP}
}
\subfigure[$q_{MP}=10^{-1}$]{
\includegraphics[width=0.3\columnwidth]{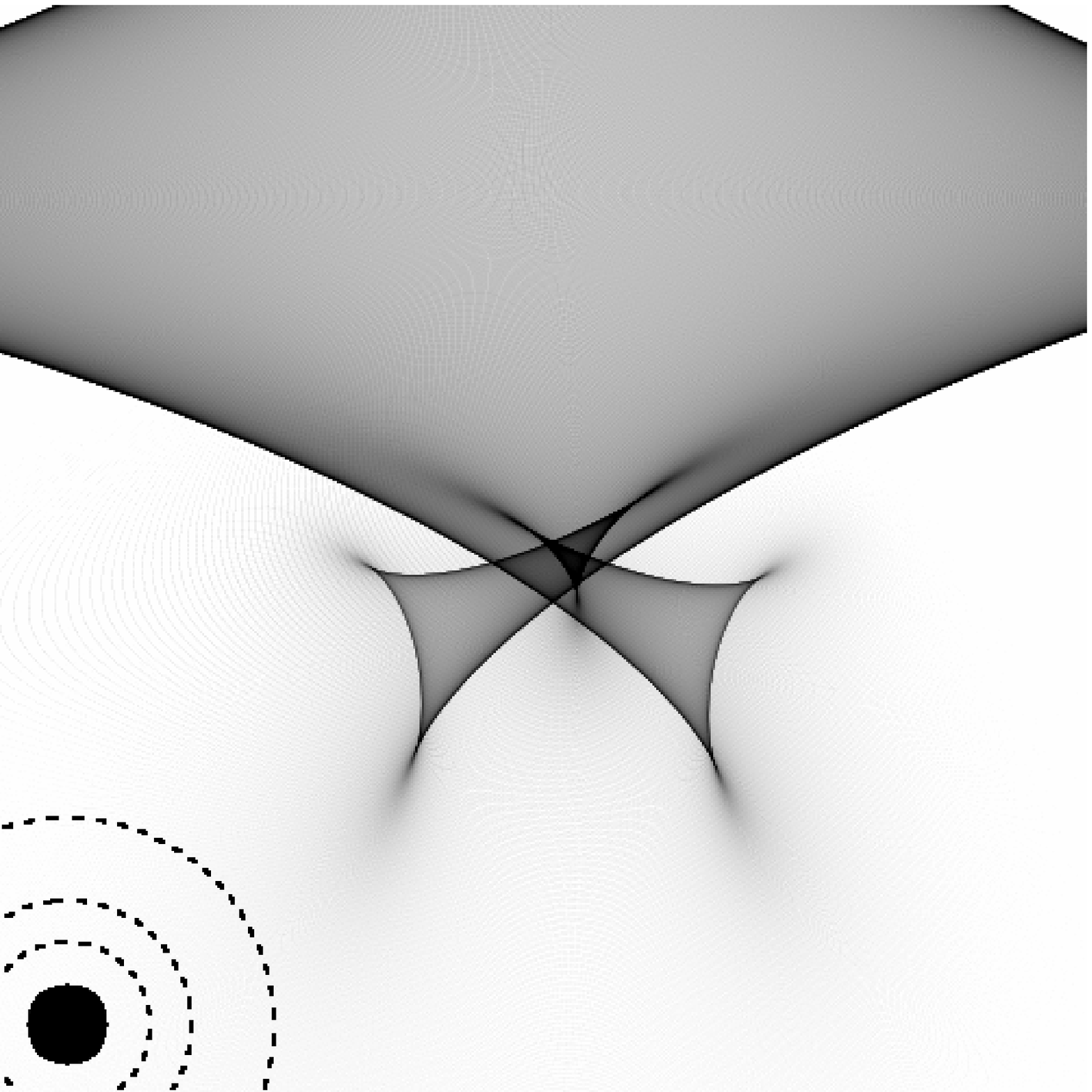}
\label{fig:sub3-caustic-q_MP}
}
\caption[Optional caption for list of figures]{Caustic topology with
  different lunar mass ratios. From left to right, the lunar mass
  ratio increases from $q_{MP}=10^{-3}$ to $10^{-1}$. The side length
  of the displayed patterns is
  $0.05\,\ensuremath{\theta_{\textit{E}}}$.  All other parameters as
  given in Table~\ref{tab:standardsummary}, in particular
  $q_{PS}=10^{-3}$ and
  $\theta_{MP}=1.0\,\ensuremath{\theta_{\textit{E}}}^P$. Source sizes
  of $1\,R\ensuremath{_{\sun}}$ to $5\,R\ensuremath{_{\sun}}$ are
  indicated in the lower left.  }
\label{fig:caustic-q_MP}
\end{figure}

To be classified as a moon, the tertiary body must have a mass
considerably smaller than the secondary. The standard case in our
examination corresponds to the Moon/Earth mass ratio of
$\frac{M_{\text{Moon}}}{M_{\text{Earth}}} = 10^{-2}$. We are generous
towards the higher mass end, and include a mass ratio of $10^{-1}$ in
our analysis, corresponding to the Charon/Pluto system.  Both examples
are singular in the solar system, but we argue that a more massive
moon is more interesting, since it can effectively stabilise the
obliquity of a planet, which is thought to be favouring the
habitability of the planet \citep{Benn2001}. At the low mass end of
our analysis we examine $q_{MP}=10^{-3}$. In
Figure~\ref{fig:caustic-q_MP}, three different caustic interferences
resulting from the three adopted mass ratios are shown.  Summarised:
For the mass ratio between moon and planet we use the three values
$q_{MP} = 10^{-3}, 10^{-2}, 10^{-1}$.

\subsection{Angular separation of planet and star $\theta_{PS}$}
\label{sec:D_PS}

The physical separation of a planet and its host star, i.e. the
semi-major axis $a_{PS}$, is not directly measurable in gravitational
lensing. Only the angular separation $\theta_{PS}$ can directly be
inferred from an observed light curve, where $0 \le \theta_{PS}D_L \le
a_{PS}$ is valid for a given physical separation $a_{PS}$ at a lens
distance $D_L$. The angular separation $\theta_{PS}$ of a binary will
evoke a certain topology of caustics, illustrated e.g. in Figure~1 of
\citet{Cassan2008b}. They gradually evolve from the close separation
case with two small triangular caustics on the far side of the star
and a small central caustic at the star position, to a large central
caustic for the intermediate case, when the planet is situated near
the stellar Einstein ring,
$\theta_{PS}=1.0\,\ensuremath{\theta_{\textit{E}}}$. This is also
called the ``resonant lensing'' case, as described in
\citet{Wambsganss1997}.  If the planet is moved further out, one
obtains a small central caustic and a larger isolated, roughly
kite-shaped planetary caustic that is elongated towards the primary
mass in the beginning, but becomes more and more symmetric and
diamond-shaped if the planet is placed further outwards.

In analogy, a ``lunar resonant lensing zone'' can be defined, when
measuring the angular separation of moon and planet in units of
planetary Einstein radii $\ensuremath{\theta_{\textit{E}}}^P =
\sqrt{q_{PS}}\,\ensuremath{\theta_{\textit{E}}}$. Viewing our solar
system as a gravitational lens system at $D_L = 6\,$kpc and with $D_S
= 8\,$kpc, Jupiter's Einstein radius in physical lengths would be
0.11\,AU in the lens plane, that is almost 10 times the semi-major
axis of Callisto. However, the smaller moons of Jupiter cover the
region out to 2 Jovian Einstein radii. If there is a moon at an
angular separation of the planet of
$0.6\,\ensuremath{\theta_{\textit{E}}}^P\lesssim \theta_{MP} \lesssim
3.0\,\ensuremath{\theta_{\textit{E}}}^P$, it will show its influence
in any of the caustics topologies. But we focus our study on the
wide-separation case for the following reasons:
Regarding the close-separation triangular caustics, we argue that the
probability to cross one of them is vanishingly small and,
furthermore, the magnification substantially decreases as they move
outwards from the star.
The intermediate or resonant caustic is well ``visible'' because it
is always located close to the peak of the single-lens curve. But all
massive bodies of a given planetary system affect the central caustic
with minor or major perturbations and deformations
\citep{Gaudi1998}. There is no reason to expect that extrasolar
systems are generally less densely populated by planets or moons than
the solar system.  Since we \emph{are} looking for very small
deviations, identifying the signature of the moon among multiple
features caused by low-mass planets and possibly more moons would be
increasingly complex.
The wide-separation case is most favourable, because the planetary
caustic is typically caused by a single planet and interactions due to
the close presence of other planets are highly unlikely
\citep{Bozza1999,Bozza2000a}. Any observed interferences can therefore
be attributed to satellites of the planet. Planets with only one
dominant moon are at least common in the solar system (Saturn \&
Titan, Neptune \& Triton, Earth \& Moon). Such a system would be the
most straightforward to detect in real data.

This makes a strong case for concentrating on the wide-separation
planetary caustic. As the standard we choose a separation of
$\theta_{PS}=1.3\,\ensuremath{\theta_{\textit{E}}}$ and also test a
scenario with $\theta_{PS}=1.4\,\ensuremath{\theta_{\textit{E}}}$. For
comparison: the maximum projected separation of Jupiter at the chosen
distances $D_S$ and $D_L$ corresponds to 1.5 solar Einstein radii.
Summarised: For the angular separation between planet and star we use
the two values $\theta_{PS} = 1.3\theta_E, 1.4\theta_E$.

\subsection{Angular separation of moon and planet $\theta_{MP}$}
\label{sec:D_MP}
As the moon by definition orbits the planet, there is an upper limit
to the semi-major axis of the moon $a_{MP}$. It must not exceed the
distance between the planet and its inner Lagrange point.  This
distance is called Hill radius and is calculated as
\begin{align*}
  r_{\text{Hill}}= a_{PS} \left(\frac{M_P}{3M_S}\right)^{\frac{1}{3}}
\end{align*}
for circular orbits, with the semi-major axis $a_{PS}$ of the
planetary orbit. This can be translated to ``lensing parameters'', but
the equation changes into the inequality 
\begin{align}\label{hillradius}
  r_{\text{Hill}} \ge \theta_{PS}D_L \left(\frac{1}{3}
    q_{PS}\right)^{\frac{1}{3}},
\end{align}
because for a physical distance $a_{PS}$, the apparent
(i.e. projected) separation lies in the range $0 \le \theta_{PS}D_L
\le a_{PS}$ depending on the inclination towards the line of sight.
Thus, the Hill radius constraint $a_{MP} \le r_{\text{Hill}}$ cannot
be regarded as a strict limit. Though, turning this argument around, a
minimum region of secured stability exists for given mass ratios and
angular planet-star separation $\theta_{PS}$.  Bodies in prograde
motion with an orbit below $0.5\, r_{\text{Hill}}$ can have long term
stability, for retrograde motion the limit is somewhat higher at
$0.75\, r_{\text{Hill}}$ (see \citet{Domingos2006} and references
therein, particularly \citet{Hunter1967}).  The region of secured
stability for our standard scenario has a radius of
\begin{align*}
  \theta_{MP} &\le 0.5\times \theta_{PS}D_L \left(\frac{1}{3}
    q_{PS}\right)^{\frac{1}{3}} = 0.045\,\ensuremath{\theta_{\textit{E}}} =
  1.53\,\ensuremath{\theta_{\textit{E}}}^{P},
\end{align*}
which leads to $a_{MP} = 0.09 \text{AU}$ in our setting. Our choice of
$\theta_{MP}$ is also motivated by a desire to have caustic
interactions between lunar and planetary caustic, a resonance, because
only under this condition the moon will give rise to light curve
features that are distinctly different from those of a low-mass
planet. If the moon is located at an angular separation from the
planet of -- depending on the position angle of the moon -- roughly
$0.6\,\ensuremath{\theta_{\textit{E}}}^P\lesssim \theta_{MP} \lesssim
3.0\,\ensuremath{\theta_{\textit{E}}}^P$, it will influence the
planetary caustic. For angular separations larger than
$3.0\,\ensuremath{\theta_{\textit{E}}}^P$, the lunar caustic barely
influences the planetary caustic, and if it is detected, it will be
detected as a distinct secondary caustic. \citet{Han2008a} already
pointed out that the lunar signal does not vanish with an increased
separation from the planet. But the lunar mass might be wrongly
identified as a second, lower-mass planet rather than a satellite of
the first planet. For comparison, the major moons of the solar system
are all at considerably smaller angular separations than
1.0$\,\ensuremath{\theta_{\textit{E}}}^P$, if observed as a
gravitational lens system from $D_L = 6\,$kpc with $D_S = 8\,$kpc,
with our Moon being most favourable at 0.4 of Earth's Einstein radii
at maximum projected separation. But as we have learnt through the
discovery of hundreds of exoplanets, we do not have to expect solar
system properties to be mirrored in exoplanetary systems.

\begin{figure}[tb]
  \centering
\subfigure[$\theta_{MP}=0.8\,\ensuremath{\theta_{\textit{E}}}^P$]{  
\includegraphics[width=0.3\columnwidth]{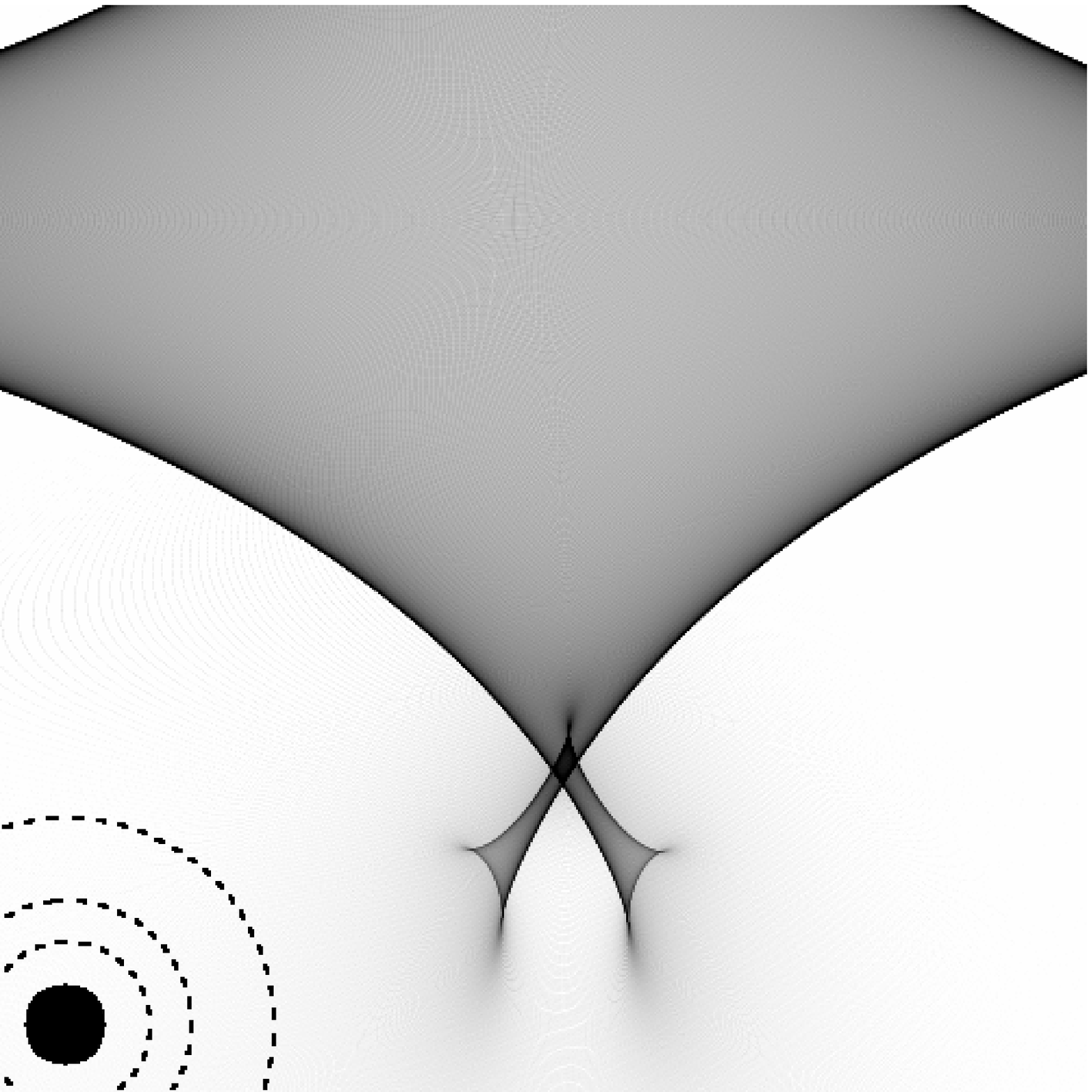}
\label{fig:sub1-caustic-d_MP}
}
\subfigure[$\theta_{MP}=1.0\,\ensuremath{\theta_{\textit{E}}}^P$]{
\includegraphics[width=0.3\columnwidth]{13844fg07b.eps}
\label{fig:sub2-caustic-d_MP}
}
\subfigure[$\theta_{MP}=1.2\,\ensuremath{\theta_{\textit{E}}}^P$]{
\includegraphics[width=0.3\columnwidth]{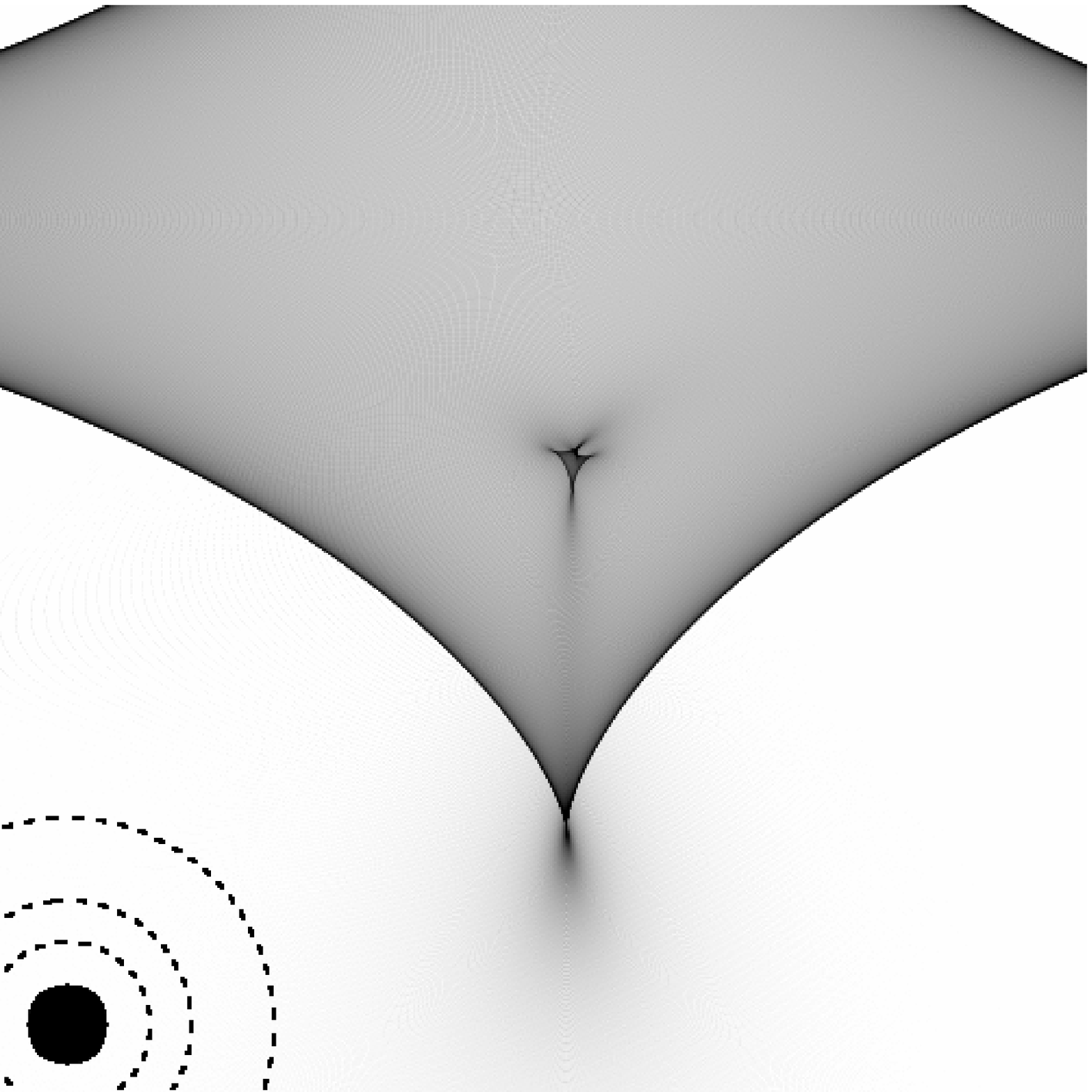}
\label{fig:sub3-caustic-d_MP}
}
\caption[Optional caption for list of figures]{Magnification patterns
  with side lengths of $0.05\,\ensuremath{\theta_{\textit{E}}}$
  showing the lunar perturbation of the planetary caustic for
  increasing angular separations of moon and planet of $\theta_{MP}=
  0.8\,\ensuremath{\theta_{\textit{E}}}^P,
  1.0\,\ensuremath{\theta_{\textit{E}}}^P,
  1.2\,\ensuremath{\theta_{\textit{E}}}^P$, while all other parameters
  are as in Table~\ref{tab:standardsummary}. The topology of
  interaction varies with the angular separation.  Source sizes of
  $1\,R\ensuremath{_{\sun}}$ to $5\,R\ensuremath{_{\sun}}$ are
  indicated in the lower left.}
\label{fig:caustic-d_MP}
\end{figure}

We focus on configurations with strong interferences of planetary and
lunar caustic, what one might call the ``resonant case of lunar
lensing'', examples of which are displayed in Figure
\ref{fig:caustic-d_MP}. Therefore, our standard case will have an
angular separation of moon and planet of one planetary Einstein
radius. 
%
%
In total, we evaluate four settings: $\theta_{MP} = 0.8
\theta_{\textit{E}}^P, 1.0 \theta_{\textit{E}}^P, 1.2
\theta_{\textit{E}}^P, 1.4 \theta_{\textit{E}}^P $, in the chosen
setting (Table~\ref{tab:standardsummary}), this corresponds to
physical projected separations from 0.5 to 0.8 AU.

\subsection{Position angle of moon $\phi$}
\label{sec:phi}
The fifth physical parameter for determining the magnification
patterns is the position angle of the moon $\phi$ with respect to
planet-star axis, as depicted in Figure~\ref{fig:mp-parameters}. It is
the only parameter that is not fixed in our standard
scenario. Instead, we evaluate this parameter in its entirety by
varying it in steps of $30\degr$ to complete a full circular orbit of
the moon around the planet. By doing this, we are aiming at getting
complete coverage of a selected mass/separation scenario.
Figure~\ref{fig:orbit} visualises how the position of the moon affects
the planetary caustic. It is not to be expected that we will ever have
an exactly frontal view of a perfectly circular orbit, but it serves
well as a first approximation. Summarised: we choose 12 equally spaced
values for the position angle $\phi$ of the moon relative to the
planet-star axis, and the results for the various parameter sets are
averaged over these 12 geometries.

\subsection{Source size $R_{\text{Source}}$}
\label{sec:source}
\begin{figure}[htb!]
  \centering 
\subfigure[$R_{\text{source}}=2R\ensuremath{_{\sun}}$.]{
  \resizebox{1.05\columnwidth}{!}{\input{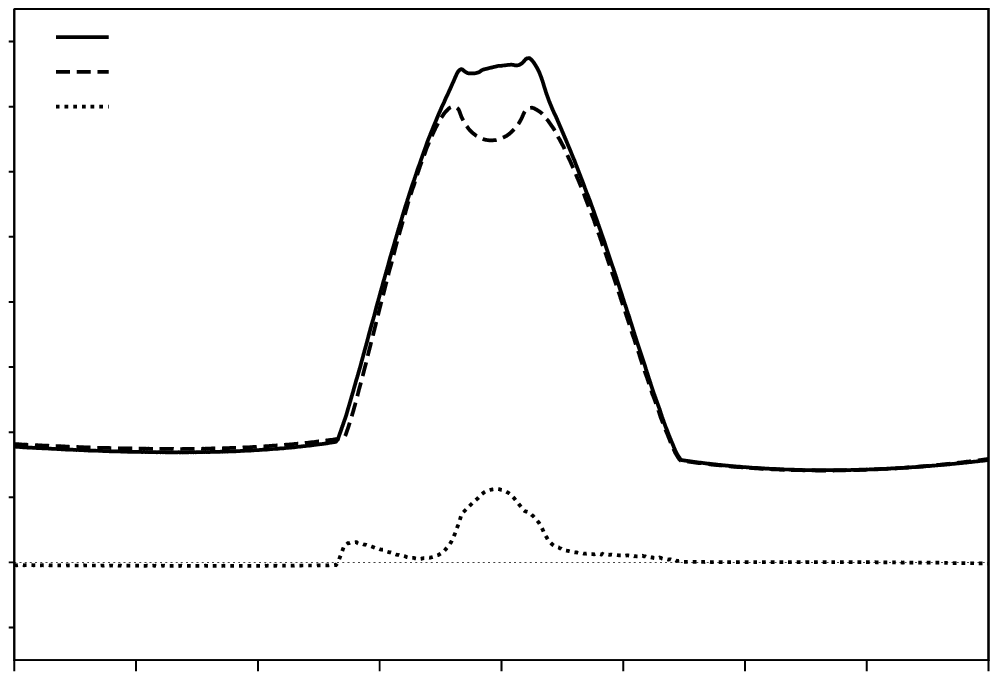}}
  \label{fig:2solarsizedsource}
}
\subfigure[$R_{\text{source}}=3R\ensuremath{_{\sun}}$.]{
\resizebox{1.05\columnwidth}{!}{\input{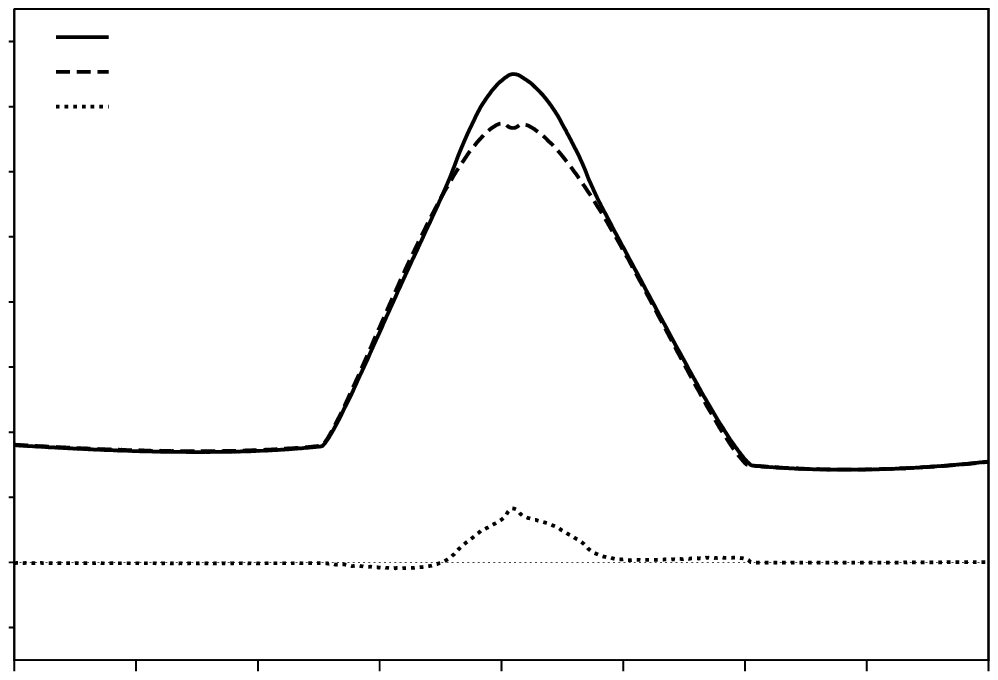}}
  \label{fig:3solarsizedsource}
}
\subfigure[$R_{\text{source}}=5R\ensuremath{_{\sun}}$.]{
  \resizebox{1.05\columnwidth}{!}{\input{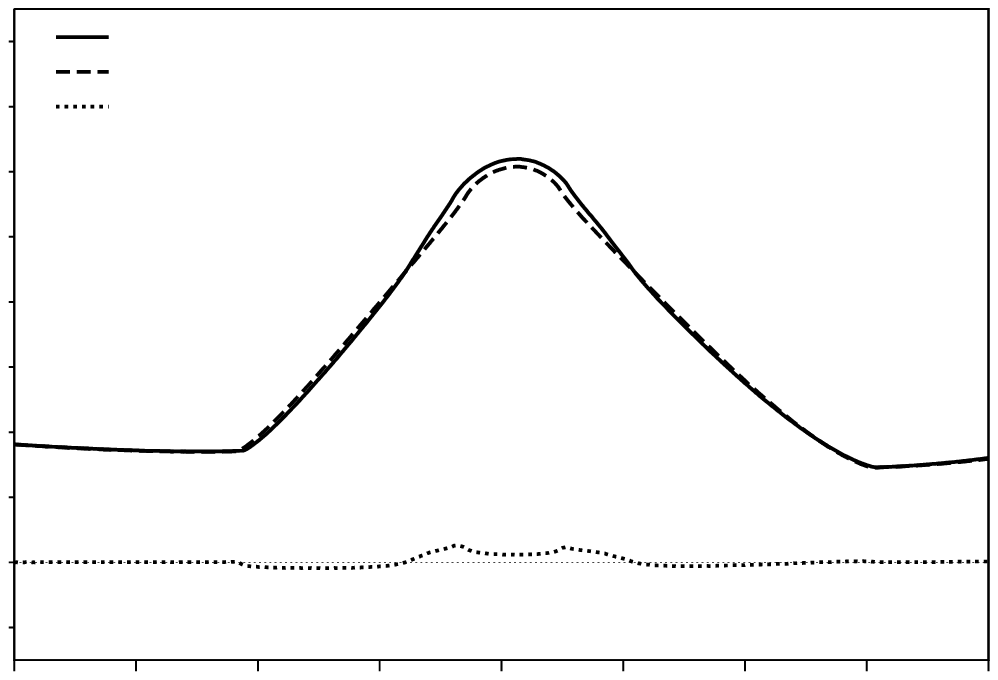}}
  \label{fig:5solarsizedsource}
}
\caption[Optional caption for list of figures]{Illustration of the
  increasing source radius, compare Figure \ref{fig:fit}:
  \subref{fig:2solarsizedsource}
  $R_{\text{source}}=2\,R\ensuremath{_{\sun}}$,
  \subref{fig:3solarsizedsource} $3\,R\ensuremath{_{\sun}}$, and
  \subref{fig:5solarsizedsource} $5\,R\ensuremath{_{\sun}}$, all other
  parameters as given in Table~\ref{tab:standardsummary}. The
  triple-lens light curve (solid) is obtained from the magnification
  pattern in Figure~\ref{fig:track-649}, the best-fit binary-lens
  light curve (dashed) from the corresponding magnification pattern
  without a moon. The difference between the two curves is plotted as
  a dotted line.}
  \label{fig:lc-sourcesizes}
\end{figure}

\begin{figure*}[bth]
\input{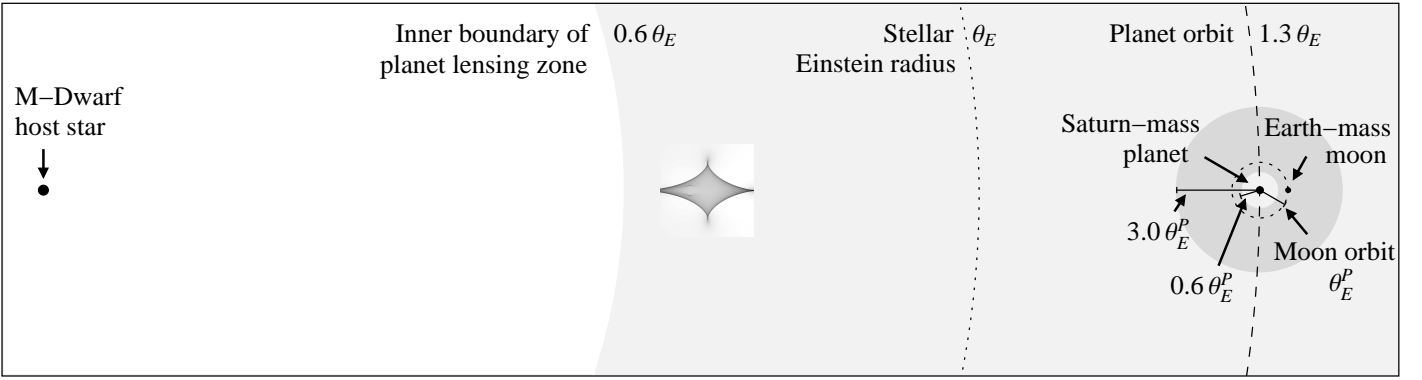}
\caption{Visualisation of the approximate lunar resonance zone
    (dark-grey ring), where planetary and lunar caustic overlap and
    interact. Also depicted is the planet lensing zone (light grey),
    which is the area of planet positions for which the planetary
    caustic(s) lie within the stellar Einstein radius. Distances and
    caustic inset are to scale. The physical scale depends not only on
    the masses of the lensing bodies, but also on the distances to the
    lens system and the background star. In our adopted standard
    scenario, cf. Table~\ref{tab:standardsummary}, the Einstein radius
    $\theta_E$ corresponds to 1.9\,AU in the lens plane. This places
    the planet at a projected orbit radius of 2.5\,AU, whereas the
    moon's orbit has a radius of 0.06\,AU (corresponding to one planetary
    Einstein radius). The lunar resonance zone covers lunar orbit radii
    from 0.04\,AU to 0.18\,AU.}
\label{fig:lensingzones}
\end{figure*}

The source size strongly influences the amplitudes of the light curve
fluctuations and the ``time resolution''. Larger sources blur out
finer caustic structures, compare Figure~\ref{fig:lc-sourcesizes}.
For our source star assumptions, we have to take into account not only
stellar abundances, but also the luminosity of a given stellar
type. Main sequence dwarfs are abundant in the Galactic bulge, but
they are faint, with apparent magnitudes of $\sim 15$ mag to 25 mag at
$D_S=8\,$kpc. In the crowded microlensing fields, they are difficult
to observe with satisfying photometry by ground-based telescopes, even
if they are lensed and magnified. Giant stars, with apparent baseline
magnitudes $\sim 13$ mag to 17 mag, are more likely to allow precise
photometric measurements from ground and, therefore, today's follow-up
observations are biased towards these source stars.

The finite source size constitutes a serious limitation to the
discovery of extrasolar moons. In fact, \citet{Han2002} stated that
detecting satellite signals in the lensing light curves will be close
to impossible, because the signals are smeared out by the severe
finite-source effect. They tested various source sizes (and
planet-moon separations) for an Earth-mass planet with a Moon-mass
satellite. They find that even for a K0-type source star
($R_{\text{Source}} = 0.8\,R\ensuremath{_{\sun}}$) any light curve
modifications caused by the moon are washed out.  \citet{Han2008a}
increased lens masses and by assuming a solar source size
$R_{\text{Source}} = R\ensuremath{_{\sun}}$, he finds that
``non-negligible satellite signals occur'' in the light curves of
planets of 10 to 300 Earth masses, ``when the planet-moon separation
is similar to or greater than the Einstein radius of the planet'' and
the moon has the mass of Earth.  We use a solar sized source as our
standard case, and present results for four more source radii: $
R_{\text{Source}} = 1 R\ensuremath{_{\sun}}, 2 R\ensuremath{_{\sun}},
3 R\ensuremath{_{\sun}}, 5 R\ensuremath{_{\sun}}, 10
R\ensuremath{_{\sun}} $.  In terms of stellar Einstein radii this
corresponds to five different source size settings from
$1.8\times10^{-3}\,\ensuremath{\theta_{\textit{E}}}$ to
$18\times10^{-3}\,\ensuremath{\theta_{\textit{E}}}$.

The ability to monitor smaller sources -- dwarfs, low-mass main
sequence stars -- with good photometric accuracy is one of the
advantages of space-based observations.  Hence a satellite mission is
the ideal tool for this aspect of microlensing, routine detections of
moons around planets can be expected with a space-based monitoring
program on a dedicated satellite.

\subsection{Sampling rate $f_\text{sampled}$}
\label{sec:sampling}

Typical exposure times for the microlensing light curves are $10$ to
$300\,$seconds at a 1.5m telescope. The frequency of observations of
an individual interesting event varies between once per 2 hours to
once every few minutes, for a single observing site only limited by
exposure time and read-out time of the instrument. One example of this
is the peak coverage of {MOA-2007-BLG-400}, see \citet{Dong2009},
where frames were taken every six minutes. Higher observing frequency
equals better coverage of the resulting light curve.  A normal
microlensing event is seen as a transient brightening that can last
for a few days to weeks or even months. A planet will alter the light
curve from fraction of a day to about a week; the effect of a low-mass
planet or a moon will only last for a few hours or shorter, see
e.g. \citet{Beaulieu2006}.  This duration is inversely proportional to
the transverse velocity $v_\perp$ with which the lens moves relative
to the line of sight or the relative proper motion $\mu_\perp$ between
the background source and the lensing system. We fix this at a typical
$v_\perp=200\,\text{km/s}$ for $D_L=6\,$kpc or
$\mu_\perp=7.0\,\text{mas/year}$.

For the simulations, a realistic, non-continuous observing rate is
mimicked by evaluating the simulated light curve at intervals that
correspond to typical frequencies, but allowing for a constant
sampling rate, which in practise is often prohibited by observing
conditions. 
We sample the triple-lens light curve 
in equally sized steps. With the assumed
constant relative velocity, this translates to equal spacing in
time. Aiming for a rate of $f_{\text{sampled}}\simeq \frac{1 \text{
    frame}}{15 \text{ minutes}}$, we choose a step size of
$6\times10^{-4}\,\ensuremath{\theta_{\textit{E}}}$.  
To see the effect
of a decreased sampling rate, we have also examined the standard
scenario (see Table~\ref{tab:standardsummary}) with sampling rates of
factors of 1.5 and 2 longer. Summarised: we use three different
sampling rates, 
$f_\text{sampled}$ = 1/15 min, 1/22 min, 1/30 min.  

\subsection{Assumed photometric uncertainty $\sigma$ }
\label{sec:sigma}

The standard error $\sigma$ of a data point in a microlensing light
curve can vary significantly. Values between $5$ and $100\,mmag$ seem
realistic, see, e.g., data plots in the recent event analyses of
\citet{Gaudi2008} or \citet{Janczak2009}.  The photometric uncertainty
directly enters the statistical evaluation of each light curve as
described in appendix \ref{sec:significance}.  We have drawn results
for an (unrealistically) broad $\sigma$-range from $0.5\,mmag$ to
$500\,mmag$. An error as small as $\sigma = 0.5\,mmag$ for a $V=12.3$
star has recently been reached with high-precision photometry of an
exoplanetary transit event at the Danish $1.54\,$m telescope at ESO La
Silla \citep{Southworth2009a}, which is one of the telescopes
presently used for Galactic microlensing monitoring observations, but
this low uncertainty cannot be transferred to standard microlensing
observations, where crowded star fields make e.g. the use of
defocussing techniques impossible.  In our discussion, we regard only
the more realistic range from $\sigma = 5$ to $100\,mmag$.  We adopt
an ideal scenario with a fixed $\sigma$-value for all sampled points
on the triple-lens light curve and set $\sigma = 20\,mmag$ to be our
standard for the comparison of different mass and separation
scenarios.  Summarised: we explore 5 values as photometric
uncertainty: $\sigma = 5$, 10, 20, 50, $100\,mmag$.

\begin{table}[htb!]
\begin{center}
  \begin{tabular}[h]{l@{\hspace{-1mm}}r l p{4.3cm}}
    \toprule
    & Parameter & Standard value & Comment\\
    \midrule
    &$D_S$ & $8\,$kpc & distance to Galactic bulge  \\
    &$D_L$ & $6\,$kpc &   \\
    &$M_{\text{Star}}$ & $0.3\,M\ensuremath{_{\sun}}$ & most abundant
    type of star  \\ 
    &$\mu_\perp$ & $7\,\text{mas/year}$ & $=v_\perp=200\,\text{km/s}$
    at $D_L=6\,$kpc  \\
    *&$q_{PS}$ & $10^{-3}$ &  Jupiter/Sun mass ratio\\
    *&$\theta_{PS}$ & $1.3\,\ensuremath{\theta_{\textit{E}}}$ &  wide
    separation caustic \\ 
    *&$q_{MP}$ & $10^{-2}$ & Moon/Earth mass ratio \\
    *&$\theta_{MP}$ & $1.0\,\ensuremath{\theta_{\textit{E}}}^P$ &
    planetary Einstein radius\\ 
    *&$R_{\text{Source}}$ & $R\ensuremath{_{\sun}}$ & brightness requirements
    vs. stellar abundance \\ 
    *&$f_{\text{sampled}}$ & $\simeq \frac{1 \text{ frame}}{15 \text{
        minutes}}$ &  high-cadence observation \\ 
    *&$\sigma$ & $20\,mmag$ & typical value in past observations \\
    \bottomrule
  \end{tabular}
\end{center}
\caption{Parameter values of our standard
  scenario. Parameters marked with an asterisk (*) are varied in our
  simulations in order to evaluate their influence on the lunar detection
  rate and to compare different triple-lens scenarios. The fixed
  parameters lead to values for the Einstein ring radius,
  $\ensuremath{\theta_{\textit{E}}} =0.32\,mas$, i.e.~1.9\,AU
    in the lens plane, and the Einstein
  time, $t_E \simeq 17\,$days. 
  The lensed system is a Saturn-mass planet at a projected
  separation of 2.5\,AU from its 0.3\,M\ensuremath{_{\sun}} M-dwarf
  host, the Earth-mass satellite orbits the planet at 0.06\,AU,
  i.e.~$0.01\,mas$ angular separation, cf. Figure~\ref{fig:lensingzones}.}
\label{tab:standardsummary}
\end{table}

\section{Results: detectability of extrasolar moons in microlensing
  light curves}
\label{sec:results}

The numerical results of our work are presented and a first
interpretation is made. We start by considering the result for a
single magnification pattern in Section~\ref{sec:output}. We 
then present results that cover specific physical triple-lens scenarios
(Section~\ref{sec:combi-results}).
We also evaluate the parameter dependence of the detectability rate, by
varying each parameter separately.

\subsection{Evaluation of a single magnification pattern}
\label{sec:output}
As described in Section~\ref{sec:method}, all magnification maps are
evaluated by taking a large sample of
unbiased\footnote{I.e. independent of lunar features, but required to
  pass close to or through the planetary caustic.} light curves and
then determining the fraction of light curves with significant lunar
signals. The results for a single
magnification map (here the pattern from
Figure~\ref{fig:lightcurvegrid}), are displayed in Table
\ref{tab:singleMP}.
\begin{table}[h]
  \begin{center}
     \begin{tabular*}{0.86\columnwidth}[h]{rrrrr}
       \toprule
       \multicolumn{2}{l}{$\sigma$ in $mmag$} &
       \multicolumn{2}{p{3.3cm}}{Proportion of significantly deviating
         light curves} & Detectability\\ 
       \midrule
       \qquad  5  && \qquad 190/228 && 83.3 \% \\
       10  && 132/228 && 57.9 \% \\
       20  &&  77/228 && 33.8 \% \\
       50  &&  39/228 && 17.1 \% \\
       100 &&   1/228 &&  0.4 \% \\
       \bottomrule
  \end{tabular*}
  \end{center}
  \caption{Detectability of the third mass in the magnification pattern of
    Figure~\ref{fig:lightcurvegrid}, for different assumed
    photometric uncertainties.}
       \label{tab:singleMP}
\end{table}
The percentages can be interpreted as the ``detection probability'' of
the moon in a random observed light curve displaying a signature of
the planetary caustic -- ignoring the possible complication to fully
characterise the signal. In this map, the detectability of the moon is
about one third, provided we have an observational uncertainty of
$\sigma = 20\,mmag$.  Obviously, with a larger photometric error, one
is less sensitive to small deviations caused by the moon.

\subsection{Results for selected scenarios}
\label{sec:combi-results}

In order to be able to make general statements about a given
planetary/lunar system, we average the results for 12 magnification
patterns representing a full lunar orbit with $\phi=0\degr,
30\degr,\ldots, 330\degr$.

\subsubsection{Variation of the photometric uncertainty}
The scenario described in Table~\ref{tab:standardsummary} is evaluated
for different photometric uncertainties by analysing the 12
magnification maps shown in Figure~\ref{fig:orbit}. The results are
displayed in Table~\ref{tab:multiMP}.
\begin{table}[h]
\begin{center}
     \begin{tabular*}{0.45\columnwidth}[h]{ l r}
       \toprule
       $\sigma$ in $mmag$ & Detectability\\
       \midrule
        \phantom{10}5  & 90.2 \% \\
        \phantom{1}10  & 59.8 \% \\
        \phantom{1}20  & 30.6 \% \\
        \phantom{1}50  & 12.8 \% \\
       100  &  1.5 \% \\
       \bottomrule
  \end{tabular*} 
\end{center}
\caption{Detectability of the moon in our standard scenario as
  a function of assumed photometric uncertainty.}
       \label{tab:multiMP}
\end{table}
For an photometric uncertainty of $\sigma=20mmag$, the result is
that about $30\,\%$ of the light curves show significant deviations
from a best-fit binary-lens light curve and display detectable signs
of the moon. We recall again our requirements:
\begin{itemize}
\item The host planet of the satellite has been detected.
\item A small source size, up to a few solar radii, is required.
\item The moon is massive compared to satellites in the
  solar system.
\item We have assumed light curves with a constant sampling rate
  of one frame every 15 minutes for about 50 to 70 hours.
\end{itemize}
As would be expected, the detection rate increases with higher
photometric sensitivity and decreases with a larger uncertainty.  At a
photometric error of $20\,mmag$ or worse, the detection of a moon
depends on whether the source trajectory passes through or close to
the lunar caustic. The finite magnification resolution of the
magnification patterns induces an uncertainty in the detectability
results that increases with smaller $\sigma$-values, but it is
negligible for the examined range.

\subsubsection{Changing the angular separation of planet and moon}
Table~\ref{tab:results-d_MP} lines up the changing detectability for
different values of the projected planet-moon separation
$\theta_{MP}$, with all other parameters as in standard scenario
(Table~\ref{tab:standardsummary}).
\begin{table}[h]
\begin{center}
     \begin{tabular*}{0.45\columnwidth}[b]{ l r}
       \toprule
       $\theta_{MP}$ in $\ensuremath{\theta_{\textit{E}}}^P$\qquad &
       Detectability\\
       \midrule
        0.8  & 29.6 \% \\
        1.0  & 30.6 \% \\
        1.2  & 30.4 \% \\
        1.4  & 28.1 \% \\
       \bottomrule
  \end{tabular*}
\end{center}
\caption{Detectability of the moon as a function of the planet-moon 
  separation for our standard scenario.} 
       \label{tab:results-d_MP}
\end{table}
With the chosen range from
$\theta_{MP}=0.8\,\ensuremath{\theta_{\textit{E}}}^P$ to
$1.4\,\ensuremath{\theta_{\textit{E}}}^P$, we are within the regime of
caustic interactions of planetary and lunar caustic. We find that the
percentage of detectability is almost constant at a $30\,\%$ level.
\citet{Han2008a} defines regions of possible satellite detections as
the region of angular moon-planet separations between the lower limit
of one planetary Einstein radius and an upper limit at the projected
Hill-radius average. From our results, we conclude that detections
with $\theta_{MP} < 1.0\,\ensuremath{\theta_{\textit{E}}}^P$ are well
possible. We have not fully probed the lower limit region, but for
$\theta_{MP} < 0.5\,\ensuremath{\theta_{\textit{E}}}^P$ the size of the
lunar caustic perturbation decreases substantially.

\subsubsection{Variation of the moon mass}
Table~\ref{tab:results-q_MP} shows the changing detectability for a
varying mass ratio $q_{MP}$ of lunar and planetary mass, with all other
parameters as in our standard scenario (Table~\ref{tab:standardsummary}).
\begin{table}[h]
\begin{center}
     \begin{tabular*}{0.45\columnwidth}[h]{ @{$}l@{$} r}
       \toprule
       q_{MP}\qquad\qquad\qquad & Detectability\\
       \midrule
        10^{-3}  &  2.4 \% \\
        10^{-2}  & 30.6 \% \\
        10^{-1}  & 85.6 \% \\
       \bottomrule
  \end{tabular*}
\end{center}
      \caption{Detectability of the moon depending on moon-planet mass
      ratio.}
       \label{tab:results-q_MP}
\end{table}
Over the range of the moon-planet mass ratio, $q_{MP}=10^{-3}$,
$10^{-2}$, and $10^{-1}$ the detection rate increases from 2.4 \% to
85.6 \%, compare also Figure~\ref{fig:caustic-q_MP}.  Not
surprisingly: the more massive the moon, the easier it is to detect.

\subsubsection{Changing the planetary mass ratio}
Table~\ref{tab:results-q_PS} lists the changing detectability for a
varying mass ratio between planet and star $q_{PS}$. 
\begin{table}[h]
\begin{center}
      \begin{tabular*}{0.45\columnwidth}[h]{ @{$}l@{$} r}
       \toprule
       q_{PS}\qquad\qquad\qquad & Detectability\\
       \midrule
        3  \times10^{-4}\qquad  & 19.6 \% \\
        1  \times10^{-3}\qquad  & 30.6 \% \\
        3  \times10^{-3}\qquad  & 42.7 \% \\
       \bottomrule
  \end{tabular*}
\end{center}
\caption{Dependence of lunar detectability as a function of the
  planet-star mass ratio, all other parameters as in our standard
  scenario.}
       \label{tab:results-q_PS}
\end{table}
We adjusted the parameters of the grid of source trajectories, so that
the grid spacing scales with the total caustic size. For a
decreasing mass of the planet, the apparent source size increases
relative to the caustic size, so finer features will be blurred out
more in the case of a smaller planet. Also the absolute change in
magnification scales with the planetary mass. As expected, the
detection efficiency is roughly proportional to the
planetary caustic size.

\subsubsection{Different sampling rates}
Table~\ref{tab:results-f} presents the changing lunar detectability for
different sampling rates.
\begin{table}[h]
\begin{center}
     \begin{tabular*}{0.75\columnwidth}[h]{p{2cm} p{1.6cm} r}
       \toprule
       Sampling step size in \ensuremath{\theta_{\textit{E}}} & 
       $f_{\text{sampled}}$ per minutes & Detectability\\
       \midrule
         $\phantom{1}6\times10^{-4}$  & 1/15 & 30.6 \% \\
         $\phantom{1}9\times10^{-4}$  & 1/22 & 26.8 \% \\
        $12\times10^{-4}$  & 1/30 & 24.9 \% \\
       \bottomrule
  \end{tabular*}
\end{center}
      \caption{Detectability of the moon depending on the sampling
        rate of the simulated light curves.}
       \label{tab:results-f}
\end{table}
A lower sampling frequency lowers the detection probability, but the
effect is less pronounced than expected.  This means it may be
favourable to monitor a larger number of planetary microlensing events
with high sampling frequency, rather than a very small number of them
with ultra-high sampling.  Our assumed sampling rates can easily be
met by follow-up observations, as they are presently performed for
anomalous Galactic microlensing events.

\subsubsection{Source size variations}

We analysed  simulated light curves of the standard scenario with five
different source star radii. They cover the range between a star the
size of our Sun and a small giant star with $R_{\text{Source}} =
10\,R\ensuremath{_{\sun}}$ at $D_S=6\,$kpc.
\begin{table}[h]
\begin{center}
     \begin{tabular*}{0.8\columnwidth}[h]{p{2.2cm} p{2.2cm}  r}
       \toprule
       $R_{\text{Source}}$ in $\ensuremath{\theta_{\textit{E}}}$\qquad &
       $R_{\text{Source}}$ in $R\ensuremath{_{\sun}}$& Detectability\\  
       \midrule
       $\phantom{1}1.8\times10^{-3}$  &$\phantom{1}$1.0 & 30.6 \% \\
       $\phantom{1}3.6\times10^{-3}$  &$\phantom{1}$2.0 & 20.9 \% \\
       $\phantom{1}5.4\times10^{-3}$  &$\phantom{1}$3.0 & 18.1 \% \\
       $\phantom{1}9.0\times10^{-3}$  &$\phantom{1}$5.0 &  4.1 \% \\
       $18.0\times10^{-3}$ & 10.0 & 0.0 \% \\
       \bottomrule
  \end{tabular*}
\end{center}
\caption{Detectability of the moon depending on the size of the
  source star, all other parameters as in our standard scenario.}
       \label{tab:results-R_Source}
\end{table}

A larger source blurs out all sharp caustic-crossing features of a
light curve, compare Figure \ref{fig:lc-sourcesizes}.  We see from our
results in Table~\ref{tab:results-R_Source} that the discovery of a
moderately massive moon (see standard scenario assumptions in
Table~\ref{tab:standardsummary}) is impossible for a source
star size of $R_{\text{Source}}=10\,R\ensuremath{_{\sun}}$ or larger within
our assumptions.

\subsubsection{Increased separation between star and planet}

For reasons given above in Section \ref{sec:D_PS}, we did not vary the
separation between star and planet significantly, but stayed in the
outer ranges of the planet lensing zone, corresponding to a
well-detectable, wide-separation planetary caustic. For
$\theta_{PS}=1.4\,\ensuremath{\theta_{\textit{E}}}$, we get a
detectability of 33.3 \%, which is similar to our standard case
($\theta_{PS}=1.3\,\ensuremath{\theta_{\textit{E}}}$) with 30.6 \%.

\section{Conclusion and outlook}
\label{sec:conclusion}

In this work we provide probabilities for the detection of moons
around extrasolar planets with gravitational lensing.  We showed that
massive extrasolar moons can principally be detected and identified
via the technique of Galactic microlensing. From our results it can be
concluded that the detection of an extrasolar moon -- under favourable
conditions -- is within close reach of available observing
technology. The unambiguous characterisation of observed lunar
features however will be challenging. The examined lens scenarios
model realistic triple-lens configurations.  An observing rate of
about one frame per 15 minutes is desirable, which is high, but well
within the range of what is now regularly performed in microlensing
follow-up observations of anomalous events.  Similarly, an
observational uncertainty of about $20\,mmag$ can be met with today's
ground-based telescopes and photometric reduction techniques for
sufficiently bright targets. Bright microlensing targets are mostly
giant stars, which is an impediment to the detection of moons: Bulge
giants with radii of order 10 $R_\odot$ or larger smooth out any lunar
caustic feature.  Therefore, in order to find moons, it is crucial to
be able to perform precise photometry on small sources with angular
sizes of the order of $10^{-3}$ Einstein radii, corresponding to a few
solar radii or smaller at a distance of 8\,kpc. This means dwarf stars
rather than giants need to be monitored in order to identify moons in
the intervening planetary systems. Some improvement in resolution can
be gained with the lucky-imaging technique on medium-sized ground
telescopes \citep{Grundahl2009}. Under very favourable circumstances,
exomoons might already be detectable from ground. However, future
space-based telescopes, such as ESA's proposed
Euclid\footnote{sci.esa.int/euclid} mission
(cf. \citet{Refregier2010}, Chapter 17) or the dedicated Microlensing
Planet Finder (MPF) mission proposed to NASA
(cf. \citet{Bennett2009b}), will surely have the potential to find
extrasolar moons through gravitational lensing.

\begin{acknowledgements}
  We thank our referee, Jean-Philippe Beaulieu, for his constructive
  comments, which helped to improve the manuscript.  CL would like to
  thank Sven Marnach for his helpful input on the significance
  test. This research has made use of NASA's Astrophysics Data System
  and the arXiv e-print service operated by Cornell University.
\end{acknowledgements}

\bibliographystyle{aa}
\bibliography{13844}

\begin{thebibliography}{49}
\expandafter\ifx\csname natexlab\endcsname\relax\def\natexlab#1{#1}\fi

\bibitem[{Beaulieu {et~al.}(2006)Beaulieu, Albrow, Bennett,
  {et~al.}}]{Beaulieu2006}
Beaulieu, J.-P., Albrow, M., Bennett, D.~P., {et~al.} 2006, Nature, 439, 437

\bibitem[{Beaulieu {et~al.}(2008)Beaulieu, Kerins, Mao, Bennett, Cassan,
  Dieters, Gaudi, Gould, Batista, Bender, Brillant, Cook, Coutures,
  Dominis-Prester, Donatowicz, Fouqu{\'{e}}, Grebel, Greenhill, Heyrovsky,
  Horne, Kubas, Marquette, Menzies, Rattenbury, Ribas, Sahu, Tsapras, Udalski,
  \& Vinter}]{Beaulieu2008}
Beaulieu, J.-P., Kerins, E., Mao, S., {et~al.} 2008, ESA white paper,
  arXiv:0808.0005v1

\bibitem[{Benn(2001)}]{Benn2001}
Benn, C.~R. 2001, Earth, Moon and Planets, 85, 61

\bibitem[{Bennett {et~al.}(2009)Bennett, Anderson, Beaulieu, Bond, Cheng, Cook,
  Friedman, Gaudi, Gould, Jenkins, Kimble, Lin, Mather, Rich, Sahu, Sumi,
  Tenerelli, Udalski, \& Yock}]{Bennett2009b}
Bennett, D.~P., Anderson, J., Beaulieu, J.-P., {et~al.} 2009, white paper,
  arXiv:0902.3000v1

\bibitem[{Bennett \& Rhie(2002)}]{Bennett2002}
Bennett, D.~P. \& Rhie, S.~H. 2002, \apj, 574, 985

\bibitem[{Bond {et~al.}(2004)Bond, Udalski, Jaroszy\'nski, Rattenbury,
  Paczy\'{n}ski, Soszy\'{n}ski, Wyrzykowski, Szyma\'{n}ski, Kubiak, Szewczyk,
  \.{Z}ebru\'{n}, Pietrzy\'{n}ski, Abe, Bennett, Eguchi, Furuta, Hearnshaw,
  Kamiya, Kilmartin, Kurata, Masuda, Matsubara, Muraki, Noda, Okajima, Sako,
  Sekiguchi, Sullivan, Sumi, Tristram, Yanagisawa, \& Yock}]{Bond2004}
Bond, I.~A., Udalski, A., Jaroszy\'nski, M., {et~al.} 2004, The Astrophysical
  Journal, 606, L155

\bibitem[{Bozza(1999)}]{Bozza1999}
Bozza, V. 1999, Astronomy and Astrophysics, 348, 311

\bibitem[{Bozza(2000)}]{Bozza2000a}
Bozza, V. 2000, \aap, 355, 423

\bibitem[{Cabrera \& Schneider(2007)}]{Cabrera2007}
Cabrera, J. \& Schneider, J. 2007, \aap, 464, 1133

\bibitem[{Cassan(2008)}]{Cassan2008b}
Cassan, A. 2008, \aap, 491, 587

\bibitem[{Domingos {et~al.}(2006)Domingos, Winter, \& Yokoyama}]{Domingos2006}
Domingos, R.~C., Winter, O.~C., \& Yokoyama, T. 2006, \mnras, 373, 1227

\bibitem[{Dominik {et~al.}(2008)Dominik, J\o{}rgensen, Horne, Tsapras, Street,
  Wyrzykowski, Hessman, Hundertmark, Rahvar, Wambsganss, Scarpetta, Bozza,
  Calchi~Novati, Mancini, Masi, Teuber, Hinse, Steele, Burgdorf, \&
  Kane}]{Dominik2008b}
Dominik, M., J\o{}rgensen, U.~G., Horne, K., {et~al.} 2008, ESA white paper,
  arXiv:0808.0004v1

\bibitem[{Dong {et~al.}(2009)Dong, Bond, Gould, Koz{\l}owski, Miyake, Gaudi,
  Bennett, Abe, Gilmore, Fukui, Furusawa, Hearnshaw, Itow, Kamiya, Kilmartin,
  Korpela, Lin, Ling, Masuda, Matsubara, Muraki, Nagaya, Ohnishi, Okumura,
  Perrott, Rattenbury, Saito, Sako, Sato, Skuljan, Sullivan, Sumi, Sweatman,
  Tristram, Yock, {The MOA Collaboration}, Bolt, Christie, DePoy, Han, Janczak,
  Lee, Mallia, McCormick, Monard, Maury, Natusch, Park, Pogge, Santallo,
  Stanek, {The $\mu$FUN Collaboration}, Udalski, Kubiak, Szyma{\'n}ski,
  Pietrzy{\'n}ski, Soszy{\'n}ski, Szewczyk, Wyrzykowski, Ulaczyk, \& {The OGLE
  Collaboration}}]{Dong2009}
Dong, S., Bond, I.~A., Gould, A.~P., {et~al.} 2009, \apj, 698, 1826

\bibitem[{Dyson {et~al.}(1920)Dyson, Eddington, \& Davidson}]{Dyson1920}
Dyson, F.~W., Eddington, A.~S., \& Davidson, C. 1920, Royal Society of London
  Philosophical Transactions Series A, 220, 291

\bibitem[{Einstein(1916)}]{Einstein1916}
Einstein, A. 1916, Annalen der Physik, 354, 769

\bibitem[{Gaudi {et~al.}(2009)Gaudi, Beaulieu, Bennett, Bond, Dong, Gould, Han,
  Park, \& Sumi}]{Gaudi2009}
Gaudi, B.~S., Beaulieu, J.-P., Bennett, D.~P., {et~al.} 2009, white paper,
  arXiv:0903.0880v1

\bibitem[{Gaudi {et~al.}(2008)Gaudi, Bennett, Udalski, Gould,
  {et~al.}}]{Gaudi2008}
Gaudi, B.~S., Bennett, D.~P., Udalski, A., Gould, A., {et~al.} 2008, Science,
  319, 927

\bibitem[{Gaudi {et~al.}(1998)Gaudi, Naber, \& Sackett}]{Gaudi1998}
Gaudi, B.~S., Naber, R.~M., \& Sackett, P.~D. 1998, \apjl, 502, L33

\bibitem[{Gaudi \& Sackett(2000)}]{Gaudi2000}
Gaudi, B.~S. \& Sackett, P.~D. 2000, \apj, 528, 56

\bibitem[{Gould(2009)}]{Gould2009c}
Gould, A.~P. 2009, in Astronomical Society of the Pacific Conference Series,
  ed. K.~Z. Stanek, Vol. 403, 86

\bibitem[{Grundahl {et~al.}(2009)Grundahl, Christensen-Dalsgaard, Kjeldsen,
  J{\o}rgensen, Arentoft, Frandsen, \& Kj{\ae}rgaard}]{Grundahl2009}
Grundahl, F., Christensen-Dalsgaard, J., Kjeldsen, H., {et~al.} 2009, The
  Stellar Observations Network Group - the Prototype, arXiv:0908.0436v1

\bibitem[{Han(2008)}]{Han2008a}
Han, C. 2008, \apj, 684, 884

\bibitem[{Han \& Han(2002)}]{Han2002}
Han, C. \& Han, W. 2002, \apj, 580, 490

\bibitem[{Holman \& Murray(2005)}]{Holman2005}
Holman, M.~J. \& Murray, N.~W. 2005, Science, 307, 1288

\bibitem[{Hunter(1967)}]{Hunter1967}
Hunter, R.~B. 1967, \mnras, 136, 245

\bibitem[{Janczak {et~al.}(2009)Janczak, Fukui, Dong, Monard, Kozlowski, Gould,
  Beaulieu, Kubas, Marquette, Sumi, Bond, Bennett, collaboration,
  collaboration, collaboration, \& collaboration}]{Janczak2009}
Janczak, J., Fukui, A., Dong, S., {et~al.} 2009, Submitted to \apj,
  arXiv:0908.0529v1

\bibitem[{Kayser {et~al.}(1986)Kayser, Refsdal, \& Stabell}]{Kayser1986}
Kayser, R., Refsdal, S., \& Stabell, R. 1986, \aap, 166, 36

\bibitem[{Kipping(2009{\natexlab{a}})}]{Kipping2009a}
Kipping, D.~M. 2009{\natexlab{a}}, \mnras, 392, 181

\bibitem[{Kipping(2009{\natexlab{b}})}]{Kipping2009b}
Kipping, D.~M. 2009{\natexlab{b}}, \mnras, 396, 1797

\bibitem[{Kipping {et~al.}(2009)Kipping, Fossey, \& Campanella}]{Kipping2009c}
Kipping, D.~M., Fossey, S.~J., \& Campanella, G. 2009, \mnras, 400, 398

\bibitem[{Lewis {et~al.}(2008)Lewis, Sackett, \& Mardling}]{Lewis2008}
Lewis, K.~M., Sackett, P.~D., \& Mardling, R.~A. 2008, \apjl, 685, L153

\bibitem[{Liebig(2009)}]{Liebig2009}
Liebig, C. 2009, Diploma thesis, Universit{\"a}t Heidelberg

\bibitem[{Mayor \& Queloz(1995)}]{Mayor1995}
Mayor, M. \& Queloz, D. 1995, \nat, 378, 355

\bibitem[{Muraki {et~al.}(1999)Muraki, Sumi, Abe, Bond, Carter, Dodd, Fujimoto,
  Hearnshaw, Honda, Jugaku, Kabe, Kato, Kobayashi, Koribalski, Kilmartin,
  Masuda, Matsubara, Nakamura, Noda, Pennycook, Rattenbury, Reid, Saito, Sato,
  Sato, Sekiguchi, Sullivan, Takeuti, Watase, Yanagisawa, Yock, \&
  Yoshizawa}]{Muraki1999}
Muraki, Y., Sumi, T., Abe, F., {et~al.} 1999, Progress of Theoretical Physics
  Supplement, 133, 233

\bibitem[{Paczy\'{n}ski(1996)}]{Paczynski1996}
Paczy\'{n}ski, B. 1996, Annual Review of Astronomy and Astrophysics, 34, 419

\bibitem[{R\'{e}fr\'{e}gier {et~al.}(2010)R\'{e}fr\'{e}gier, Amara, Kitching,
  Rassat, Scaramella, Weller, \& {Euclid Imaging Consortium}}]{Refregier2010}
R\'{e}fr\'{e}gier, A., Amara, A., Kitching, T.~D., {et~al.} 2010,
  arXiv:1001.0061v1

\bibitem[{Rhie(1997)}]{Rhie1997}
Rhie, S.~H. 1997, The Astrophysical Journal, 484, 63

\bibitem[{Rhie(2002)}]{Rhie2002}
Rhie, S.~H. 2002, How Cumbersome is a Tenth Order Polynomial?: The Case of
  Gravitational Triple Lens Equation, arXiv:astro-ph/0202294v1

\bibitem[{Sartoretti \& Schneider(1999)}]{Sartoretti1999}
Sartoretti, P. \& Schneider, J. 1999, \aaps, 134, 553

\bibitem[{Scharf(2006)}]{Scharf2006}
Scharf, C.~A. 2006, \apj, 648, 1196

\bibitem[{Schneider {et~al.}(2006)Schneider, Kochanek, \&
  Wambsganss}]{Schneider2006}
Schneider, P., Kochanek, C.~S., \& Wambsganss, J. 2006, Gravitational Lensing:
  Strong, Weak and Micro, Saas-Fee Advanced Courses 33 (Springer-Verlag Berlin
  Heidelberg)

\bibitem[{Schneider \& Wei\ss(1986)}]{Schneider1986}
Schneider, P. \& Wei\ss, A. 1986, \aap, 164, 237

\bibitem[{Southworth {et~al.}(2009)Southworth, Hinse, J{\o}rgensen, Dominik,
  Ricci, Burgdorf, Hornstrup, Wheatley, Anguita, Bozza, Calchi~Novati,
  Harps{\o}e, Kj{\ae}rgaard, Liebig, Mancini, Masi, Mathiasen, Rahvar,
  Scarpetta, Snodgrass, Surdej, Th\"{o}ne, \& Zub}]{Southworth2009a}
Southworth, J., Hinse, T.~C., J{\o}rgensen, U.~G., {et~al.} 2009, \mnras, 396,
  1023

\bibitem[{Udalski {et~al.}(1992)Udalski, Szyma\'nski, Ka{\l}u\.{z}ny, Kubiak,
  \& Mateo}]{Udalski1992}
Udalski, A., Szyma\'nski, M., Ka{\l}u\.{z}ny, J., Kubiak, M., \& Mateo, M.
  1992, Acta Astronomica, 42, 253

\bibitem[{Wambsganss(1990)}]{Wambsganss1990}
Wambsganss, J. 1990, PhD thesis, Ludwig-Maximilians-Universit\"at M\"unchen

\bibitem[{Wambsganss(1997)}]{Wambsganss1997}
Wambsganss, J. 1997, \mnras, 284, 172

\bibitem[{Wambsganss(1999)}]{Wambsganss1999a}
Wambsganss, J. 1999, Journal of Computational and Applied Mathematics, 109, 353

\bibitem[{Williams \& Knacke(2004)}]{Williams2004}
Williams, D.~M. \& Knacke, R.~F. 2004, Astrobiology, 4, 400

\bibitem[{Witt(1990)}]{Witt1990}
Witt, H.~J. 1990, \aap, 236, 311

\end{thebibliography}

\begin{appendix}

\section{Significance of deviation}
\label{sec:significance}

This section explains the method that we use to determine the
significance of deviation between the two simulated light curves of
the triple lens and the best-fit binary lens model.

When we want to decide whether real observational data is better
described with a triple-lens or with a binary-lens model, we can fit
both models to the data and use the $\chi^2$-test to assess the
significance of the deviation between the data and the models and pick
the model with the better fit. 

We do not have observational data, but numerically computed light
curves. If we observed one of the simulated triple-lens systems, we
would expect the data points to have a Gaussian distribution around
the theoretical values. We could produce artificial data by randomly
scattering the simulated triple-lens light curve data points around
their theoretical values and then fit a binary-lens model and the
triple-lens model to this artificial data set, as described in
Section~\ref{sec:curvecomparison}. This approach allows us to use the
usual $\chi^2$-test, but is computationally intensive, because the
models have to be fitted numerous times to get a statistically sound
sample of $\chi^2$-values.  In our approach, we directly calculate the
$\chi^2$-value which is the expectation value of the above method.

We hypothesise, every simulated point on our triple-lens light curve
is in fact the mean $\mu_i^t$ of the distribution of a random variable
$X_i$ that is distributed according to the Gaussian probability
density function $f_i(x_i)$ with a standard deviation of
$\sigma_i^t$. $(\sigma_i^t)^2$ is the variance of the
distribution.\footnote{We recall the definitions of mean and
  variance. For the random variable $X$ with probability density
  function $f(x)$, they are in general
\begin{align*}
  \mu &= \int_{-\infty}^{\infty} x\;f(x)\;dx = \langle X\rangle
  \quad\text{ and} \\ 
  \sigma^2 &= \int_{-\infty}^{\infty} (x-\mu)^2\;f(x)\;dx= \langle
  (X-\mu)^2 \rangle,
\end{align*}
where we also introduce the notation $\langle X\rangle$ for the
expectation value of $X$.}

We now want to examine whether the $X_i$ could be described equally
well with a binary-lens model $\mu_i^b$. We introduce the new
$\chi^2$-distributed\footnote{The $\chi^2$-distribution can be applied
  to the sum of $n$ \emph{independent, normally distributed random
    variables} $Z_i$ with the \emph{mean} $\mu_i=0$ and the
  \emph{standard deviation} $\sigma_i=1$ for all $i$,
 \begin{align*}
   \chi^2 \sim \sum_{i=1}^n Z_i^2.
 \end{align*}
 See any standard book with an introduction to calculus of
 probability.} 
random variable
\begin{align*}
  Q^2 = \sum_{i=1}^n\left(\frac{X_i - \mu_i^b}{\sigma_i^b}\right)^2.
\end{align*}
At this step, we could simulate data $X_i$ in order to find a somewhat
representative, randomly drawn value of $Q^2$, but instead we simply
calculate what the mean value of all possible $Q^2$ would be. We use
the definitions, to find
\begin{align*}
  \langle Q^2\rangle &= \left\langle  \sum_{i=1}^n \left(\frac{X_i -
        \mu_i^b}{\sigma_i^b}\right)^2\right\rangle %
  = \sum_{i=1}^n\frac{1}{(\sigma_i^b)^2} \left\langle (X_i -
    \mu_i^b)^2\right\rangle\\
\end{align*}
Here, we keep in mind that, in general, $\langle x^2\rangle \neq
\langle x\rangle ^2$.  Using the parameters $\mu_i^t$ and $\sigma_i^t$
of the distribution $f_i(x_i)$, we reduce the equation by calculating
\begin{align*}
  \langle Q^2\rangle &= \sum_{i=1}^n
  \frac{1}{(\sigma_i^b)^2}\int_{-\infty}^{\infty} (x_i -
  \mu_i^b)^2 f_i(x_i) dx_i.\\
  &= \sum_{i=1}^n \frac{1}{(\sigma_i^b)^2}\int_{-\infty}^{\infty} (x_i
  -
  \mu_i^t + \mu_i^t - \mu_i^b)^2 f_i(x_i) dx_i\\
  &=\sum_{i=1}^n \frac{1}{(\sigma_i^b)^2} \left((\sigma_i^t)^2
    +(\mu_i^t
    - \mu_i^b)^2\right).\\
\end{align*}
We used the definitions of $\mu_i^t$ and $(\sigma_i^t)^2$ and the
property of the probability density function $\int f_i(x_i)dx_i=1$. In
our case $\sigma_i^t = \sigma_i^b$ holds true without loss of
accuracy, and indeed, we simplify $\sigma_i = \sigma$ for all $i$,
that is we assume that the photometric uncertainty is the same for all
data points, and argue that this does not pose a problem as long as
$\sigma$ is chosen to match the maximum photometric uncertainty. So we
can further reduce to
\begin{align*}
  \langle Q^2\rangle = \sum_{i=1}^n(1+\frac{1}{\sigma^2}(\mu_i^t -
  \mu_i^b)^2)= n+\sum_{i=1}^n\left(\frac{\mu_i^t -
      \mu_i^b}{\sigma}\right)^2.
\end{align*}
This is the mean of all $Q^2$ possibly resulting, when comparing the
simulated binary-lens light curve to randomly scattered triple-lens
light curve points. It shall be our measure of deviation. Fortunately,
we can provide all remaining parameters:
\begin{description}[$(\mu_i$]
\item [$n$] is the number of degrees of freedom, equal to the number of
  compared data points.
\item [$\sigma$] is the assumed standard deviation or photometric
  uncertainty of observations.
\item [$(\mu_i^t - \mu_i^b)^2$] equals the difference between the two
  compared light curves squared, which we already use for our least
  square fit.
\end{description}
\begin{figure}[tbhp]
  \centering
  \resizebox{1.04\columnwidth}{!}{\input{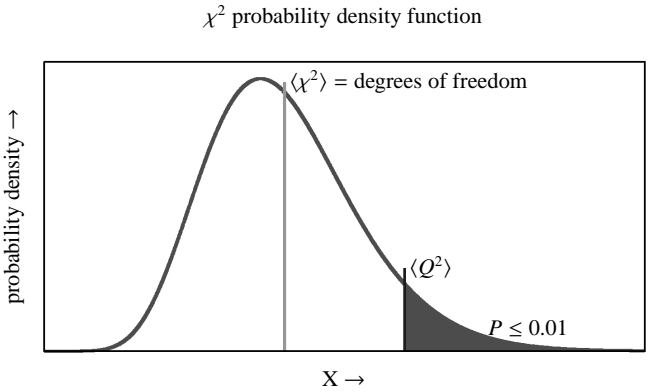}}
  \caption{The $\chi^2$ probability density function plotted for
    illustration only. For increasing degrees of freedom the central
    limit theorem takes effect and the curve will very closely
    resemble a Gaussian distribution. If $P(X\ge\langle Q^2 \rangle)
    \le 1\%$, i.e. if the probability $P$ for a $\chi^2$-distributed
    random variable $X$ to be larger or as large as the mean of $Q^2$
    is less than $0.01$, we consider the deviation between the two
    compared curves to be significant.}
  \label{fig:chi2-pdf}
\end{figure}
Now we have to decide whether the deviation between our two examined
light curves is significant.  By taking $\langle Q^2\rangle$ as
$\chi^2$, one can evaluate the cumulative distribution function $F_n(z)
= \int^z_{-\infty} f(x)dx$ of the $\chi^2$-distribution - we relied on
the GNU Scientific
Library\footnote{\url{http://www.gnu.org/software/gsl/}} - to find the
corresponding probability for any given $\chi^2$-distributed random
variable $X$ to be as large as or larger than $\langle Q^2\rangle$,
\begin{align*}
  P(X \ge \langle Q^2\rangle) = 1 - F_n(\langle Q^2\rangle).
\end{align*}
This probability is just the integral of the $\chi^2$ probability
density function to the right of $\langle Q^2\rangle$ as illustrated
in Figure~\ref{fig:chi2-pdf}. If this probability is very small,
$\langle Q^2\rangle$ is outside the expected range for a
$\chi^2$-distributed random variable $Q^2$. In that case, $Q^2$ is
obviously not $\chi^2$-distributed with $n$ degrees of freedom, so we
must conclude that there is a significant deviation between the triple
lens model light curve and the binary-lens model light curve.

We say we have a significant deviation between our two curves, if the
mean value $\langle Q^2\rangle$ is so high that the probability $P$
for any random variable $X$ being larger or equally large is less than
$1\%$.  We interpret this to say, only if the probability for a given
triple lens light curve with independent, normally distributed data
points to be random fluctuation of the compared binary-lens light
curve is less than $1\%$, we consider it to be principally detectable.

There are two known sources for overestimating the detection rate
coming with our method. First, systematic errors are not accounted
for.  A possible solution could be to add a further term to $\langle
Q^2\rangle$ as in $\langle Q_{new}^2 \rangle = \langle Q^2 \rangle -
n\left(\frac{\sigma_{sys}}{\sigma}\right)^2$, where $\sigma_{sys}$ can
be assumed to lie in the few percent region for real data.  Secondly,
we avoid refitting the binary-lens model light curve and always
compare to the one we got as best-fit to the exact (simulated)
triple-lens model light curve. This overestimation effect vanishes for
$n\rightarrow\infty$ and is negligible for a sufficiently large $n$.
The high sampling frequency we use, ensures that the number of data
points is always large enough ($n>250$ in all cases).

While we cannot exactly quantify the uncertainty of the results, we
know that they pose strict upper limits for the lunar detectability in
the various scenarios. This knowledge would enable us to infer a
tentative census of extrasolar moons, once the first successful
microlensing detections have been achieved.

\end{appendix}

\end{document}